\documentclass{bioinfoX}
\copyrightyear{2010}
\pubyear{2010}

\usepackage{amsmath}
\usepackage{multirow}
\usepackage[dvips]{epsfig}

\begin{document}
\firstpage{1}

\title[A simple and fast method to determine the parameters for fuzzy c--means cluster validation]{A simple and fast
method to determine the parameters for fuzzy c--means cluster validation}
\author{Veit Schw\"ammle$^1$\footnote{to whom
correspondence should be addressed}, and Ole N\o{}rregaard Jensen$^1$}
\address{$^1$Department of Biochemistry and Molecular Biology, University of Southern Denmark,Campusvej 55,DK-5230
Odense M,Denmark}

\history{Received on XXXXX; revised on XXXXX; accepted on XXXXX}

\editor{Associate Editor: XXXXXXX}

\maketitle

\begin{abstract}

\section{Motivation:}
Fuzzy c-means clustering is widely used to identify cluster structures in high-dimensional data sets, such as
those obtained in DNA microarray and quantitative proteomics experiments. One of its
main limitations is the lack of a computationally fast method to determine the two parameters \emph{fuzzifier} and
\emph{cluster number}. 
Wrong parameter values may either lead to the inclusion of purely random fluctuations in the results or
ignore potentially important data. The optimal solution has parameter values for which the
clustering does not yield any results for a purely random data set but which detects cluster formation with maximum
resolution on the edge of randomness. 
\section{Results:}
Estimation of the optimal parameter values is achieved by evaluation of the
results of the clustering procedure applied to randomized data sets. In this case, the optimal value of the
fuzzifier follows common rules that depend only on
the main properties of the data set. Taking the dimension of the set and the number of objects as input values instead
of evaluating the entire data set allows us to propose a functional relationship determining its value directly.
This result speaks strongly against setting the fuzzifier equal to 2 as typically done in many previous studies.
Validation indices are generally used for the 
estimation of the optimal number of clusters. A comparison shows that the minimum
distance between the centroids provides results that are at least equivalent or better than those obtained by other computationally more expensive
indices.
\section{Contact:} \href{veits@bmb.sdu.dk}{veits@bmb.sdu.dk}
\end{abstract}

\section{Introduction}
\label{introduction}

\noindent
New experimental techniques and protocols allow experiments with high resolution and thus lead to the
production of large amounts of data. In turn, these data sets demand effective machine-learning techniques for extraction of information. Among them, the recognition of patterns in noisy data still remains a challenge. The aim
is to merge
the outstanding ability of the human brain to detect patterns in extremely noisy data with the power
of computer-based automation. Cluster validation allows to group high-dimensional data points that exhibit similar properties and so to discover a possible functional relationship within subsets of data.

Different approaches to the problem of cluster validation exist, such as hierarchical clustering~\citep{Eisen98},
k-means
clustering~\citep{Tavazoie99}, and self-organizing maps~\citep{Tamayo99}. Noise or background signals in collected data
normally come from
different sources, such as intrinsic noise from variation within the sample and noise coming from the
experimental equipment. An appropriate method to find clusters in this kind of data is based on fuzzy c-means
clustering~\citep{Dunn73,Bezdek81} due to its robustness to noise~\citep{Hanai2006}. Although this method has been
modified and extended many times (for an overview see~\citet{Doring2006}),
the original procedure~\citep{Bezdek81} remains the most commonly used to date.

In contrast to k-means clustering, the fuzzy c-means procedure involves an additional parameter, generally called the
fuzzifier.
A data point (e.g. a gene or protein, from now on called an object) is not directly assigned to a cluster but is allowed
to obtain \emph{fuzzy} memberships to all
clusters. This makes it possible to decrease the effect of data objects that do not belong to one particular cluster,
for example objects located between overlapping clusters or objects resulting from background noise.
These objects, by having rather distributed membership values, now have a low influence in the calculation of the
cluster center positions. Hence, with the introduction of this new parameter, the cluster validation becomes much more
efficient in dealing with noisy data. The value of the fuzzifier defines the maximum fuzziness or noise in the
data set.
Whereas the k-means clustering procedure always finds clusters independently on the extent of
noise in the data, the fuzzy method allows first to adapt the method to the present amount of noise and second to
avoid erroneous detection of clusters generated by random patterns. Therefore, the challenge consists in
determining an appropriate value of the fuzzifier.

Usually, the value of the fuzzifier is set equal to two~\citep{Pal95,Babuska98,Hoeppner99}. This may be considered a
compromise between an a
priori assumption of a certain amount of fuzziness in the data set and the advantage of avoiding a time--consuming calculation of its value.  
However, by carefully adjusting the fuzzifier, it should be possible to optimize the algorithm to
take into account the characteristic noise present in the data set. We are interested in having maximal sensitivity to observe
barely detectable cluster structures combined with a low probability of assigning clusters originating from random
fluctuations.

Nowadays, cluster validation is in widespread use for the analysis of microarray data to discover genes with similar
expression changes. 
Recently, large data sets from quantitative proteomics, for instance measuring the peptide/protein
expression by means of mass spectrometry, became available.
These samples are usually low-dimensional, i.e. they have a small number of data points per peptide/protein. As will be
also shown in this work, low dimensionality may lead to difficulties to discard noisy patterns without
loosing all information in the data set. 

To our knowledge, only few methods exist to determine an
optimal value of the fuzzifier. 
In \citet{Dembele2003}, the fuzzifier is obtained with an empirical method calculating the coefficient of variation
of a function of the distances between all objects of the entire data set. Another approach searches for a minimal
fuzzifier value for which the cluster analysis of the randomized data set produces no meaningful results, by comparing a
modified partition coefficient for different values of both parameters~\citep{Futschik2005}. The calculations in these
two methods imply operations on the entire data set and becomes computationally expensive for large data
sets. 

Here, we introduce a method to determine the value of the fuzzifier without using the current data set. For high-dimensional data
sets, the fuzzifier value depends directly on the dimension of the data set and its number of objects, and so avoids
processing of the complete data set. For low-dimensional data sets with small numbers of objects,
we
were able to considerably reduce the search space for finding the optimal value of the fuzzifier. This improvement
helps to choose the right parameter set and to save computational
time when processing large data sets. 

Our study shows that the optimal fuzzifier generally takes values far from the frequently used value 2. We focused mainly on the clustering of biological data coming from
gene expression analysis of microarray data or from protein quantifications.
However, the present method can be applied to any data set for which one wants to detect clusters of non-random origin.

In the following section the algorithm of fuzzy c-means clustering is introduced and
the concept to avoid random cluster detections is explained. We present a simplified model
showing a strong dependence of the fuzziness on the main properties of the data set and confirm this result by
evaluating randomized artificial data sets. We distinguish between valid and invalid cluster validations by looking at
the minimal distances between the found centroids. This relationship is quantified by fitting a
mathematical function to the results for the minimum centroid distance. 

Finally, we determine the second parameter of the cluster validation, the number of clusters. Different
validation indices
are compared for artificial and real data sets.

\section{Data set and algorithm}
\noindent
Clustering algorithms are often used to analyze a large number of objects, for example genes in
microarray data,
each containing a number of values obtained at different experimental conditions. In other terms, the data set
consists of $N$ object vectors of $D$ dimensions (experimental conditions), and thus an optimal framework contains $N
\times
D$ experimental values. The aim is to group these objects into clusters with similar behaviors. 

Missing values can be replaced for example by the average of the existing
values for the object. 
In gene expression data and in quantitative proteomics data, the values of each
object represent
only a relative quantity to be compared to the other values of the object. Therefore, the focus is on fold-changes and
not on absolute value changes (a 2-fold, i.e. 200\%, increase has the same weight as a 2-fold decrease, 50\%). In this
case, the values are transformed by taking their logarithm before the data is to be evaluated.
Each object is normalized to have values with mean 0 and standard deviation 1. 

The fuzzy c-means clustering for a given parameter set ${c,m}$ -- the number of clusters and the fuzzifier --
corresponds to minimizing the objective function, 
\begin{equation}
 \label{eq:fcm}
J(c,m) = \sum \limits_{k=1}^c \sum \limits_{i=1}^N (u_{ki})^m |\mathbf{x_i}-\mathbf{c_k}|^2 ~,
\end{equation}
where we used Euclidean metrics for the distances between centroids $\mathbf{c_k}$ and objects $\mathbf{x_i}$. Here,
$u_{ki}$ denotes the membership value of object $i$ to the cluster $k$, satisfying the following criteria,
\begin{equation}
 \label{eq:fcm2}
\sum \limits_{k=1}^c u_{ki} = 1~;~0<u_{ki}<1~.
\end{equation}
The following iteration scheme allows the calculation of the centroids and the membership values by solving
\begin{equation}
  \label{eq:it1}
  \mathbf{c}_k = \frac{\sum \limits_{i=1}^N (u_{ki})^m \mathbf{x}_i}{\sum \limits_{i=1}^N (u_{ki})^m} 
\end{equation}
for all $k$ and afterwards obtaining the membership values through
\begin{equation}
  \label{eq:it2}
  u_{ki} =  \frac{1}{\sum \limits_{s=1}^c
\left(\frac{|\mathbf{x}_i-\mathbf{c}_k|^2}{|\mathbf{x}_i-\mathbf{c}_s|^2}\right)^{\frac{1}{m-1}}}~.
\end{equation}

A large fuzzifier value suppresses outliers in data sets, i.e. the larger $m$, the more clusters share their
objects and vice versa. For $m\rightarrow 1$, the method becomes equivalent to k-means clustering whereas for
$m\rightarrow
\infty$ all data objects have identical membership to each cluster. 

We minimize the objective function $J(c,m)$ by carrying out 100 iterations of Eqs.~(\ref{eq:it1}) and ~(\ref{eq:it2}).
The application of Eqs.~(\ref{eq:it1}-\ref{eq:it2}) converges to a solution that might be trapped in a local
minimum, requiring the user to repeat the minimization procedure several times with different initial conditions.
In order to be able to carry
out a vast parameter study, we limited the evaluation to 5-10 performances per data set and parameter set, taking the
performance  corresponding to the best clustering result, i.e. the one with the smallest final value of the objective
function.

The final classification of a data set into different clusters in fuzzy clustering is not as clear as in the case of k-means clustering
where each object is assigned to exactly one cluster. In
fuzzy c-means clustering, each object belongs to each cluster, to the degree given by the membership value. The
centroid, i.e. the center of a cluster, corresponds therefore to the center of all objects of the data set, each
contributing with
its own membership value. As a consequence, we need to define a threshold that defines whether an object belongs to a
certain cluster. Ideally, this threshold is
set to $1/2$. Hence, due to the limitation of Eq.~(\ref{eq:fcm2}), each object belongs to maximally one cluster.
A non-empty cluster with at least one object having a membership value greater than $1/2$ is called a \emph{hard
cluster}. 

The number of hard clusters $c_{\text{final}}$ found in the cluster validation can be smaller than the number of
previously defined clusters, $c$.
Therefore we can define the case $c_{\text{final}} < c$ to be a case of \emph{no solution} for the application of
the cluster validation. In other words, a cluster validation leading to at least one empty cluster will not be
considered as a valid performance. 

By distinguishing cases for which the cluster validation gives a valid result and cases of invalid
results it is possible to identify parameter regions where the algorithm identifies clusters that may
result from random fluctuations.
As example, take a data set and its randomized counterpart. We now fix $c$ and compare the results of the
clustering for increasing fuzzifier values, $m$. At
$m=1$,
the cluster validation is equivalent to k-means clustering, assigning exactly one cluster to each object and the
no-solution case does not exist. The clustering of both the original and the randomized data set will give $c$ valid
clusters. By increasing the value of the fuzzifier, the membership values of outliers become more distributed between
the clusters whereas objects pertaining to real clusters get their largest membership value decreased only slightly.
Each cluster looses object members with membership values larger than $1/2$ and the total number of objects that are
assigned to a cluster as hard members decreases. As the
objects of a randomized data set are distributed almost
homogeneously in cluster space, the clustering algorithm stops to detect a total of $c$
hard clusters above a certain threshold of the fuzzifier. When further increasing $m$, also the objects in the original
data set will have their largest membership values fall below $1/2$ and so the clustering of the original
data will stop to produce valid results above another threshold of $m$.  The parameter region between these two
thresholds
is of particular interest. Within this region, only the clustering of the original data set produces valid results and
thus the found clusters can be understood to correspond to non-random object groupings. Precisely, we prefer to take an
as
low as possible value of the fuzzifier, combining minimal fuzziness and maximal cluster
recognition.
The procedure presented in the next sections shows how to obtain a minimal value of $m$ that still does not give a valid
solution for the clustering of the randomized data set.
A data set having the same threshold for both the clustering of the original
set and the randomized one should be discarded as it is too noisy. However, we will see that the value of the fuzziness
increases strongly for low-dimensional data sets and thus a compromise between accepting 
clusters with members of noisy origin and low detection of patterns must be found. 



\section{Arguments for a functional relationship between the fuzzifier and the
  data set structure}
\label{sec:arg}
\noindent
A strong relationship between the fuzzifier and the basic properties of the data set can be demonstrated by
means of a simplified model system. With increasing
dimension, clusters are less likely to be found in a completely random data set. In order to illustrate this dependency
mathematically, one might reduce the system to a binary $D$-dimensional object space, i.e. $x_{id} \in \{-1,1\}$. Let us
now look at a cluster that contains an accumulation of objects at a given point in object space. E.g., for a
purely random object, the probability to have $\mathbf{x}_{i} = (1,1,...,1)$ is given by $2^{-D}$. Furthermore, the
probability to have half of all objects of the data set with this particular value equals to,
\begin{equation}
 \label{eq:prob}
\binom{N}{\frac{N}{2}} 2^{-D \frac{N}{2}} \left( 1-2^{-D} \right)^{\frac{N}{2}} \approx \sqrt{\frac{2}{\pi
N}}2^{N(1-\frac{D}{2})} \left( 1- 2^{-D} \right)^{\frac{N}{2}}~,
\end{equation}
\begin{table}[t!]
\processtable{Summary of the parameters. \label{table:par}}
{\begin{tabular}{ll}\toprule
  Parameters of the clustering   & Parameters of the artificial data set \\\midrule
$m$:   fuzzifier & $N$:  number of objects\\
&   $D$:  number of dimensions of an object \\
$c$:  number of clusters & $M$:  number of Gaussian--distributed clusters\\
& $N_O$:  number of data points per cluster\\
& $w$:  standard deviation of Gaussian\\\botrule
\end{tabular}}{}
\end{table}
\noindent
where we used the Stirling approximation. For $2^{-D}\ll 1$, the right side of Eq.~(\ref{eq:prob}) might be approximated
by $2^{N (1-\frac{D}{2})} \sqrt{2/(\pi N)}$. 
Hence, the probability for a well defined cluster decreases exponentially with respect to the dimension
of the data set, and slightly slower for a increasing number of objects in the set.
As a consequence, the clustering parameter value $m$ being a measure for the fuzziness of the system should
follow these tendencies at least qualitatively. This finding argues strongly against an application of the
fuzzy algorithm by merely using $m=2$. We will show that the simplified  model predicts the dependencies on
both quantities in the right way.

An extensive evaluation of the clustering procedure is carried out using artificially generated data sets as input. Each
object corresponds to a random point generated out of
$D$-dimensional Gaussian distributions with standard deviation $w$. The data set consists of $M$ Gaussian--distributed
clusters with each having $N_O$
objects, leading to a total of $N=N_O \times M$ objects in the set. Each Gaussian is centered at a random position in
object space,
having
coordinates between 0 and 10 for each dimension. An optimal cluster validation should identify $c=M$ as best
solution. The parameters of the fuzzy c-means algorithm and the parameters of the artificial data set are
summarized in Table~\ref{table:par}.

\begin{figure*}[!tb]
 \centering
 \includegraphics[angle=270,width=0.21\textwidth]{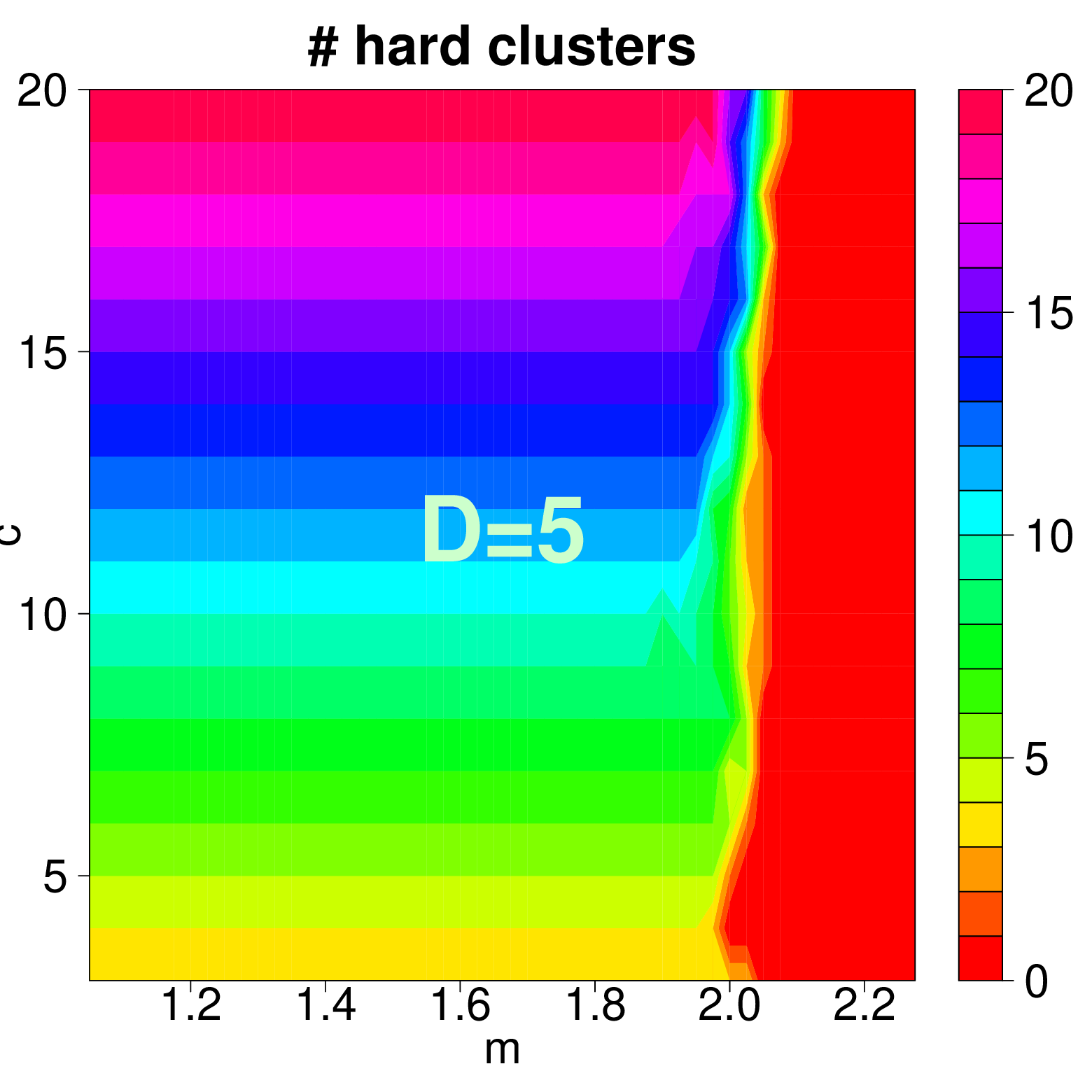}
 \includegraphics[angle=270,width=0.21\textwidth]{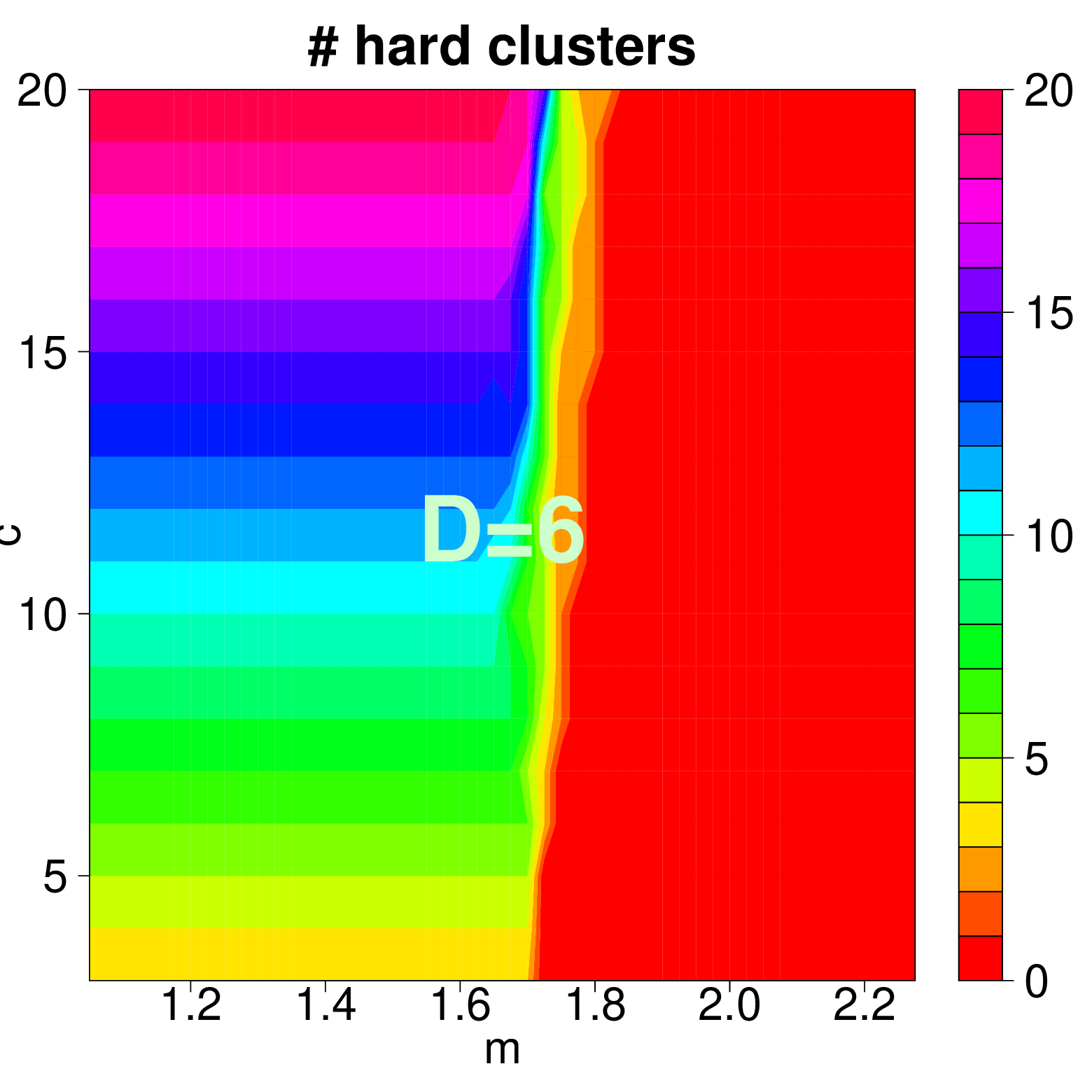}
 \includegraphics[angle=270,width=0.21\textwidth]{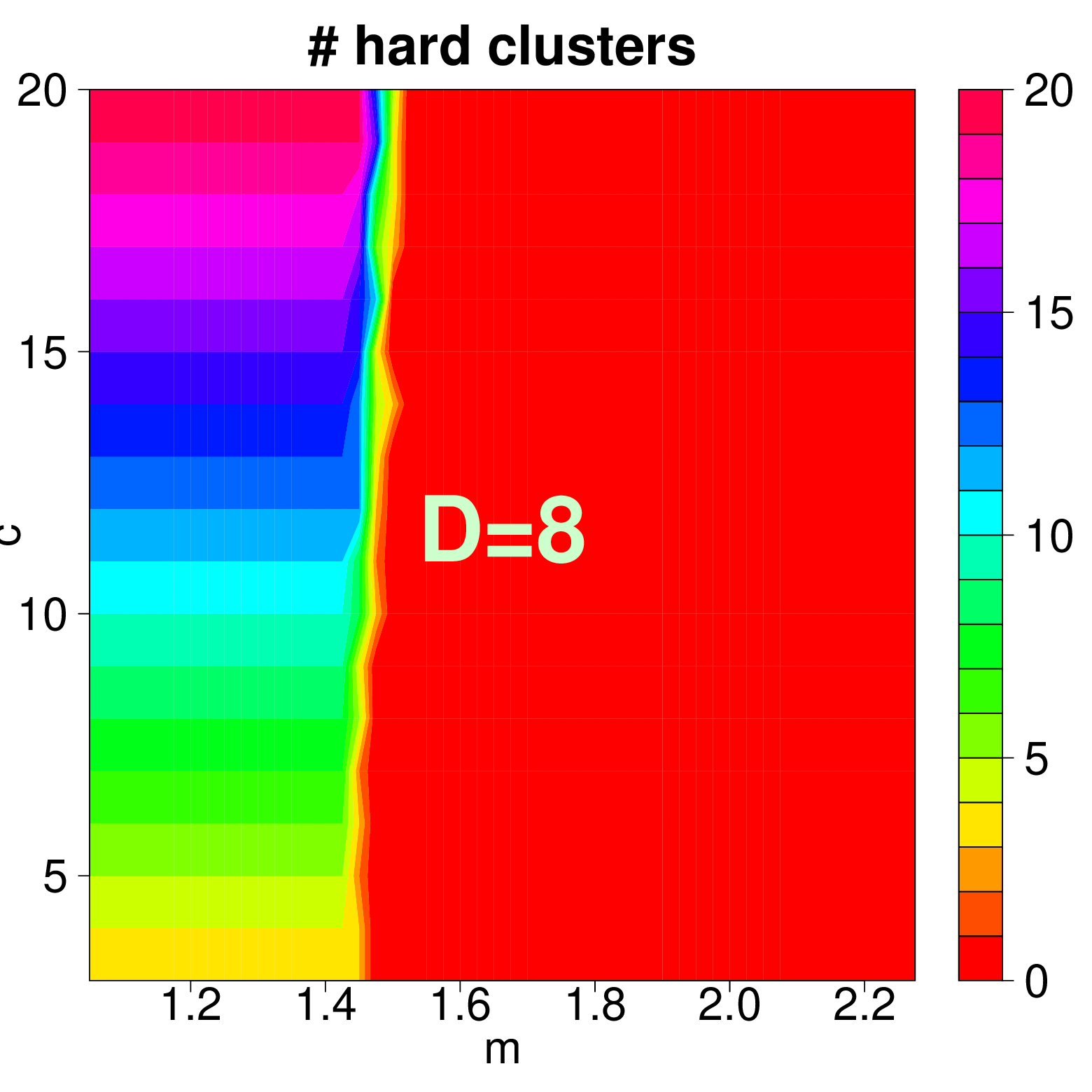}
 \includegraphics[angle=270,width=0.21\textwidth]{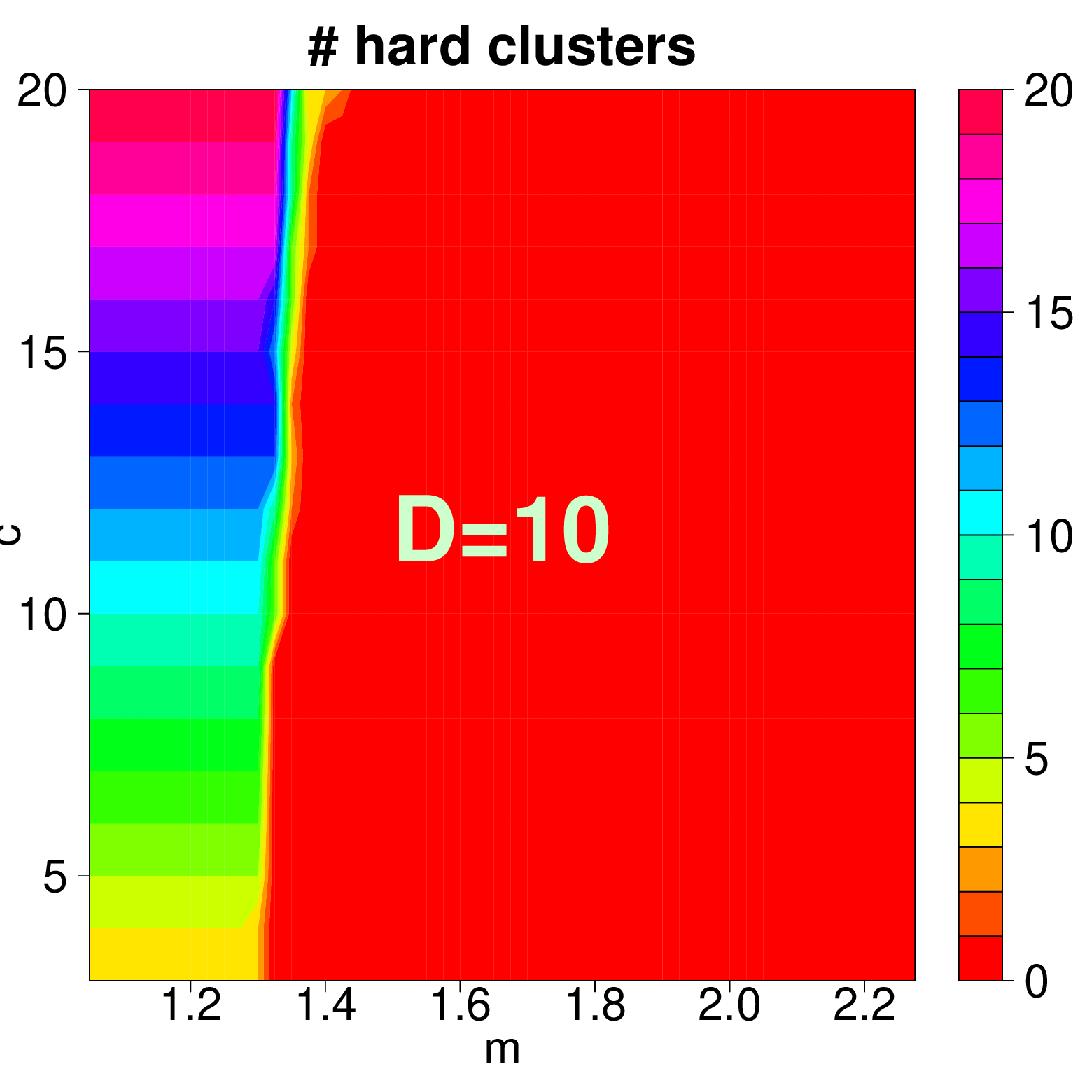}
 \includegraphics[angle=270,width=0.21\textwidth]{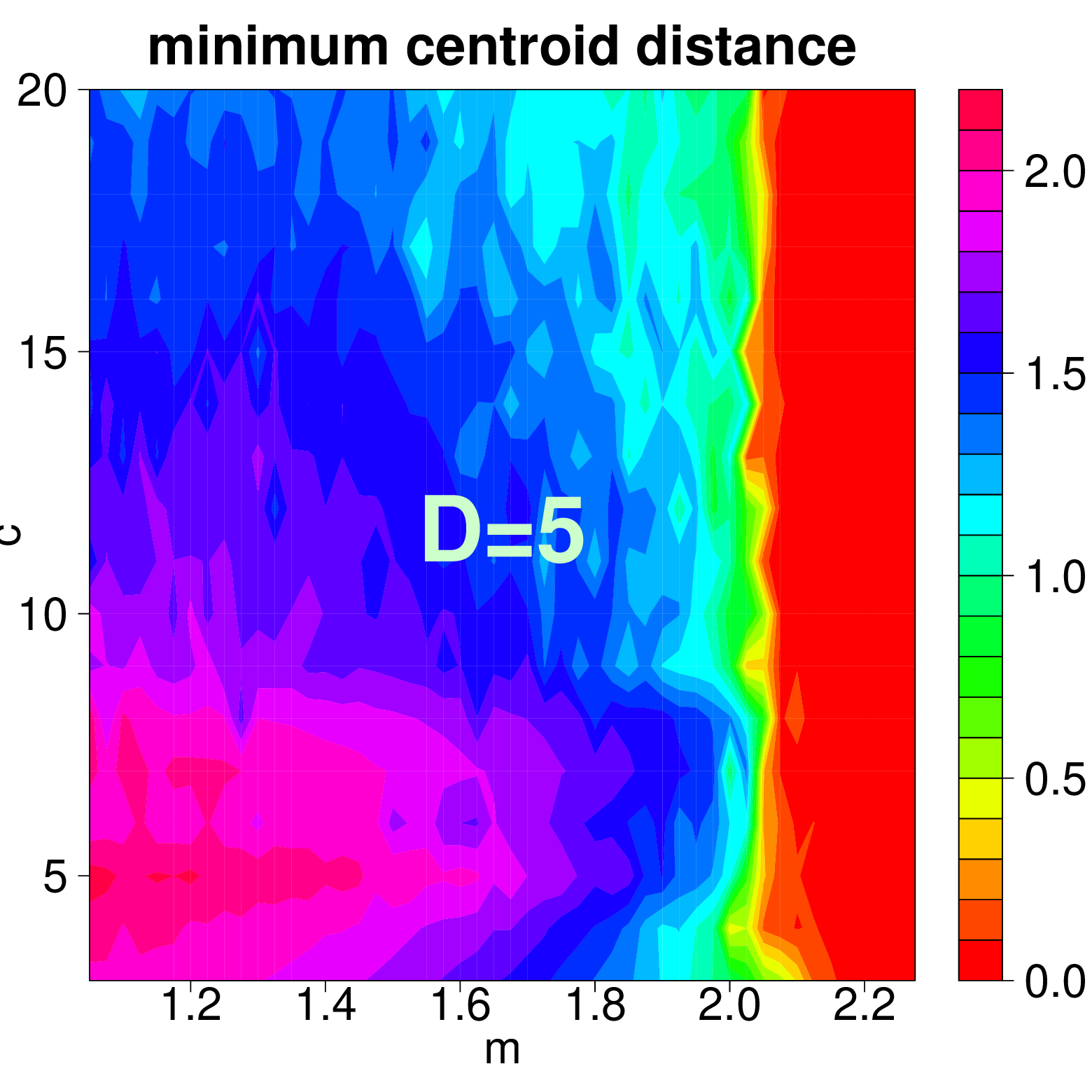}
 \includegraphics[angle=270,width=0.21\textwidth]{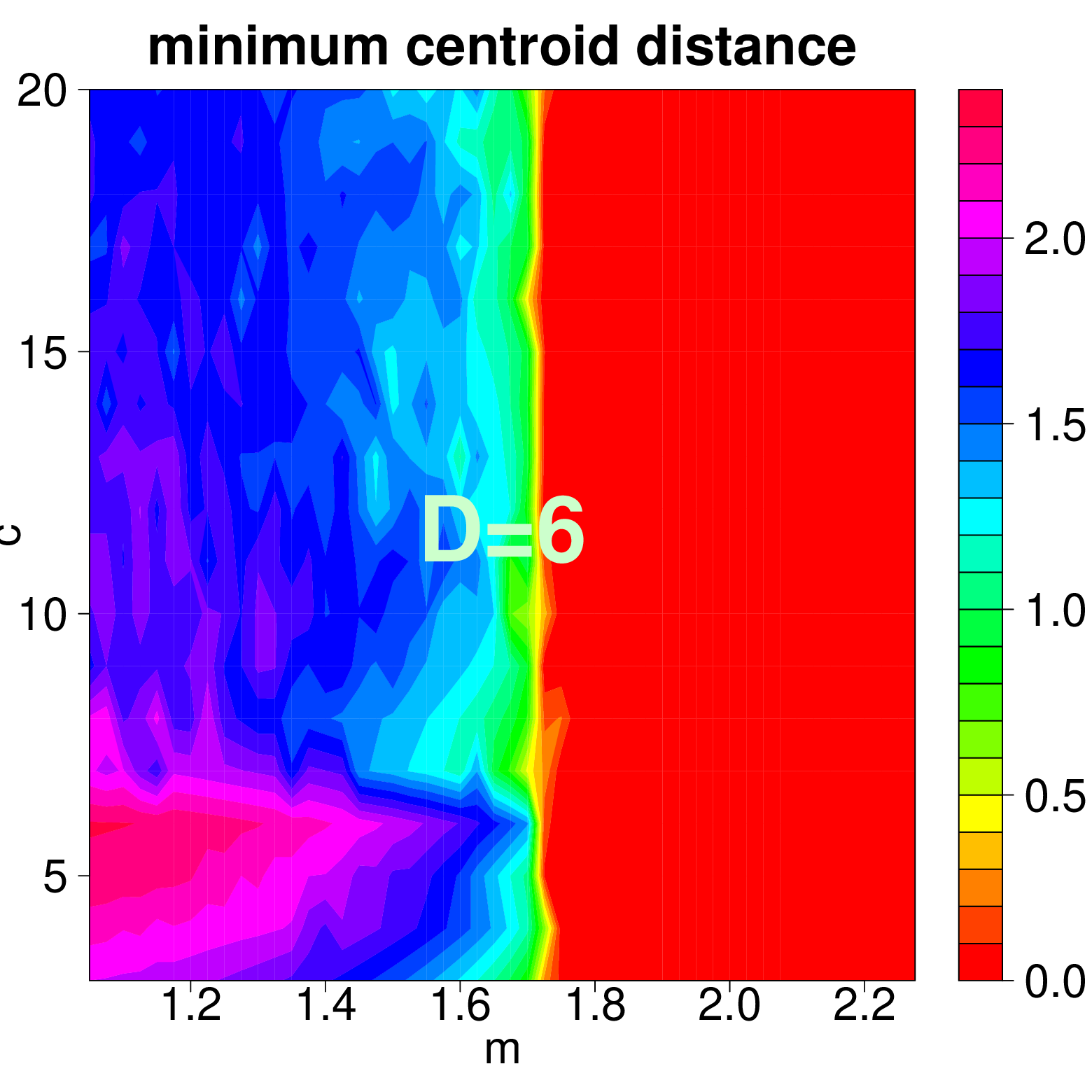}
 \includegraphics[angle=270,width=0.21\textwidth]{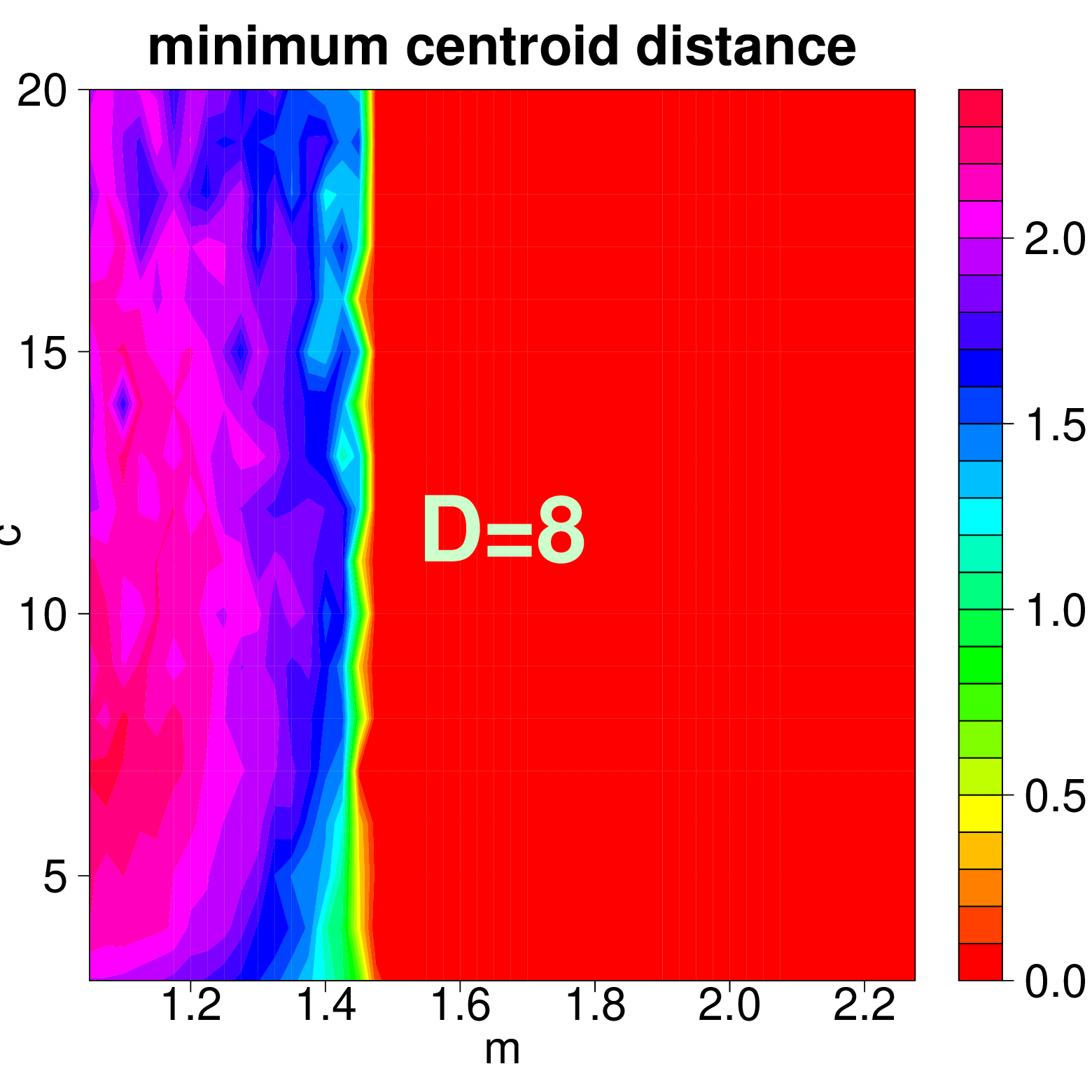}
 \includegraphics[angle=270,width=0.21\textwidth]{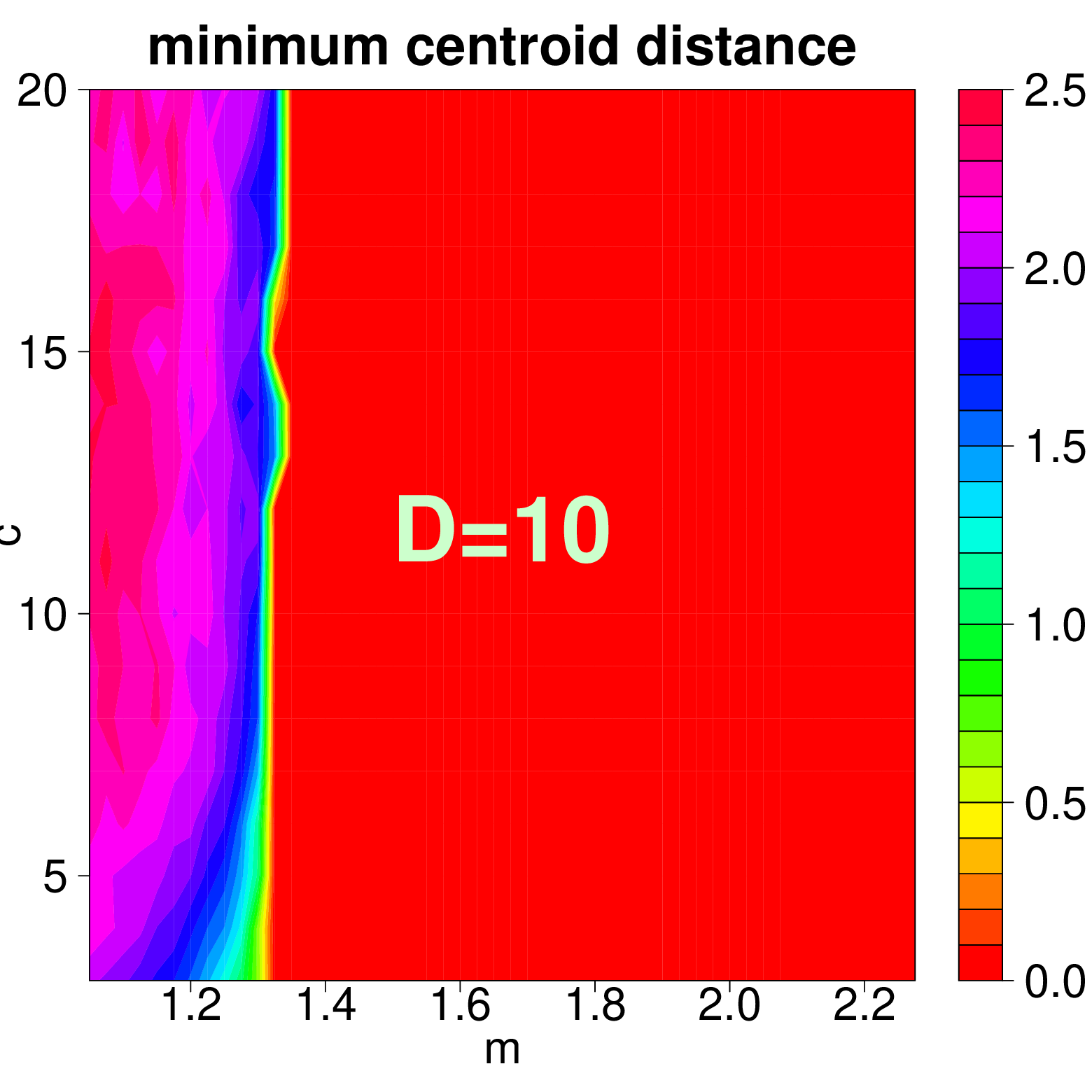}
\caption{Showing the number of hard clusters  and the minimum centroid distances of
randomized data sets for different values of $m$ (horizontal axis) and $c$ (vertical axis). The object points
are Gaussian-distributed with standard deviation $w=1$, dimensions 5,6,8 and 10, and were
randomized afterwards. There are 500
objects per data set. The threshold of $m$ where
the number of hard clusters becomes smaller than $c$ and the minimum centroid distance approaches zero does
not vary
significantly for different $c$. Moreover, the $m$-position of the threshold is the same for both measures within the
same
dimension.
}
\label{fig:c3}
\end{figure*}

A first step to find an optimal value of the fuzzifier consists in applying the clustering procedure to randomized
data sets. We generate these sets by random permutations of the values of each object. A
threshold for the fuzzifier value $m$ is reached
as soon as the clustering procedure does not provide any
valid solution for the randomized set. This corresponds to the case
where the number of hard clusters
is smaller than the value of the parameter $c$. However, another criteria allows a more accurate
estimation. We will refuse a clustering solution having two centroids that coincide, i.e. their mutual
distance falls below some predefined value. 

Fig.~\ref{fig:c3} shows both, the number of hard clusters as well as the minimum centroid distance for
different realizations of artificial data sets. There is a sudden decay to zero of both quantities when increasing
the fuzzifier. Three important conclusions can be made from the results depicted in Fig.~\ref{fig:c3}:
First, the position of the decay of the minimal centroid distances coincides with the one  where the number
of hard clusters 
changes from $c$ to $c-1$. Apparently, a cluster without any membership values over the $1/2$ limit (an empty cluster)
has always its centroid coincide with the centroid of one of the hard clusters. We could not
find any
mathematical explanation for this behavior, but our analysis shows that this relation seems to be a general
characteristics of the fuzzy c-means algorithm. Second, the minimum centroid distance decay occurs at almost exactly the
same 
value of $m$ over the entire range of $c$. This seems to be typical behavior in randomized data sets. 
Third, the $m$-position of the decay decreases for higher dimensions of the data set. High-dimensional data sets have
a structure where random clusters are less likely as already illustrated with the simplified model
presented above.
We will take the minimum centroid distance to measure the
$m$-value of the threshold in the following analysis, which is from now on denoted $m_t$.

%
%
%

%
\begin{figure*}[htb]
 \centering
 \includegraphics[angle=270,width=0.21\textwidth]{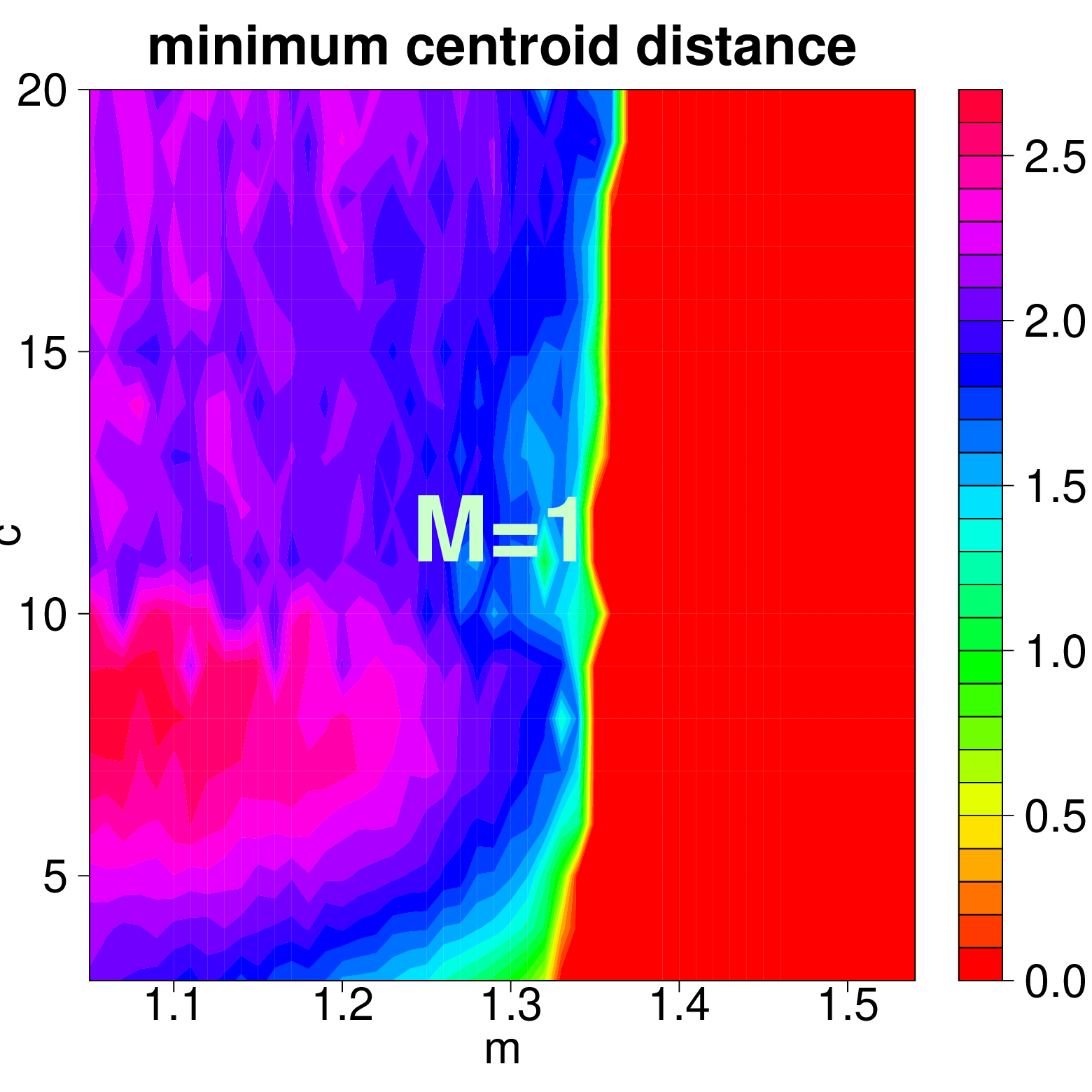}
 \includegraphics[angle=270,width=0.21\textwidth]{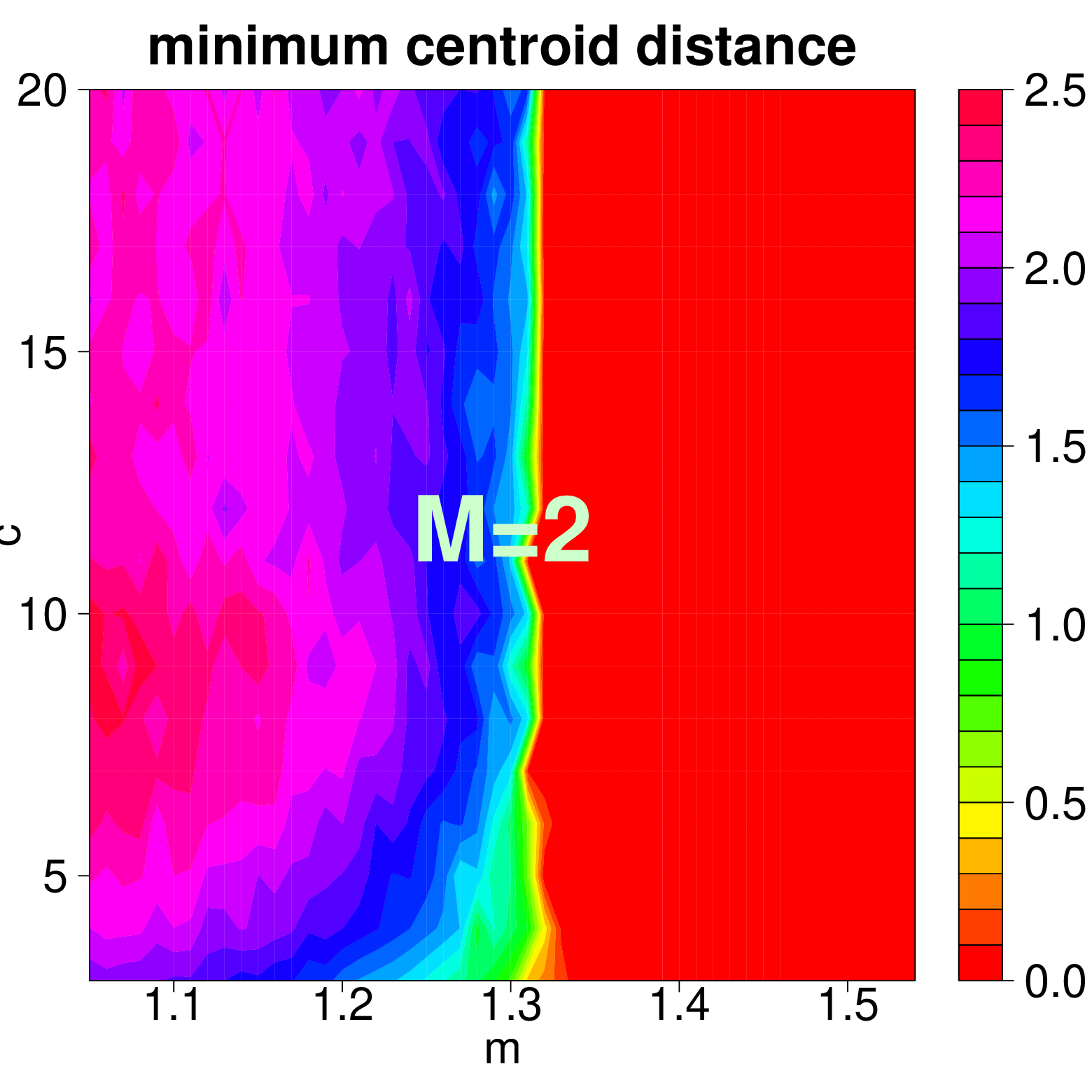}
 \includegraphics[angle=270,width=0.21\textwidth]{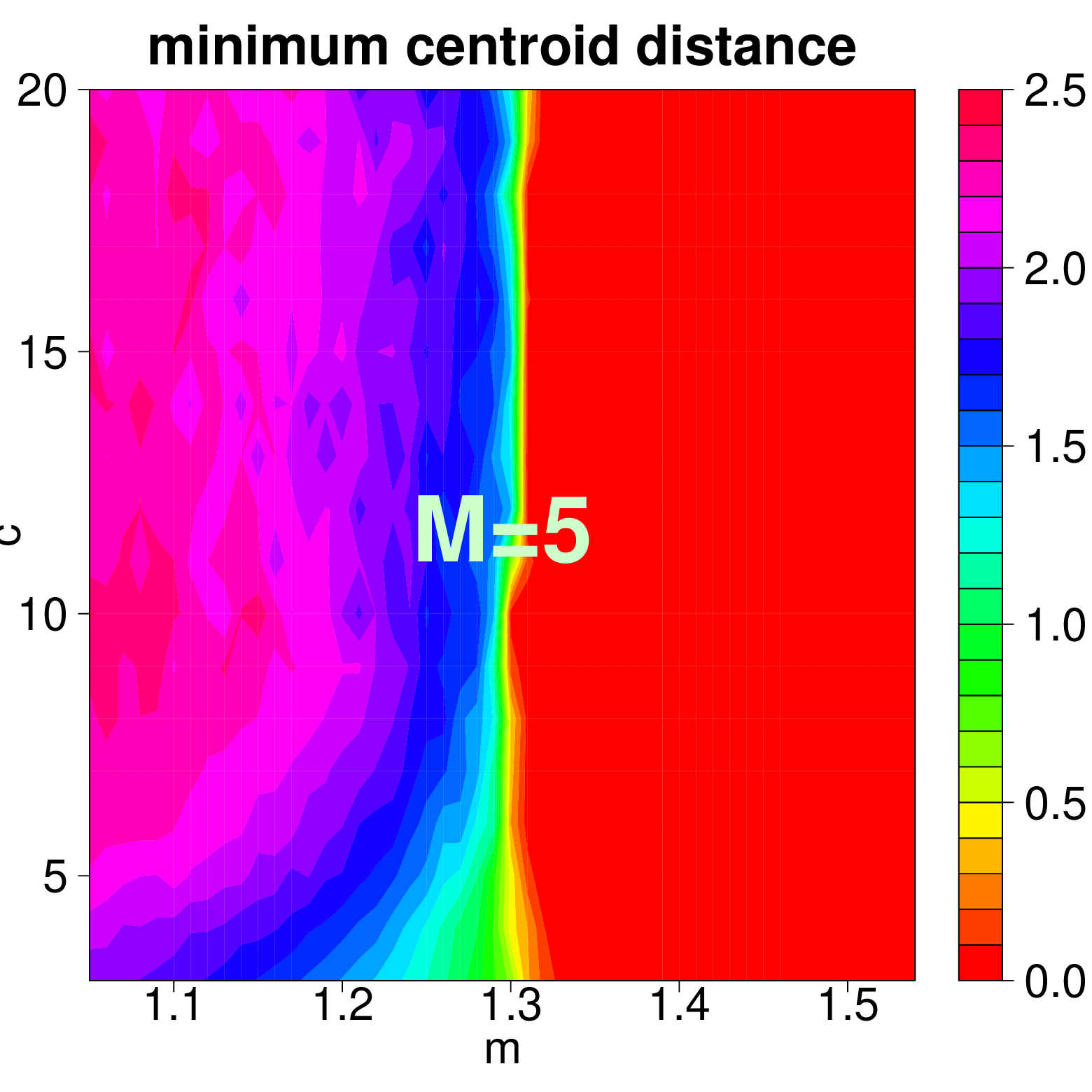}
 \includegraphics[angle=270,width=0.21\textwidth]{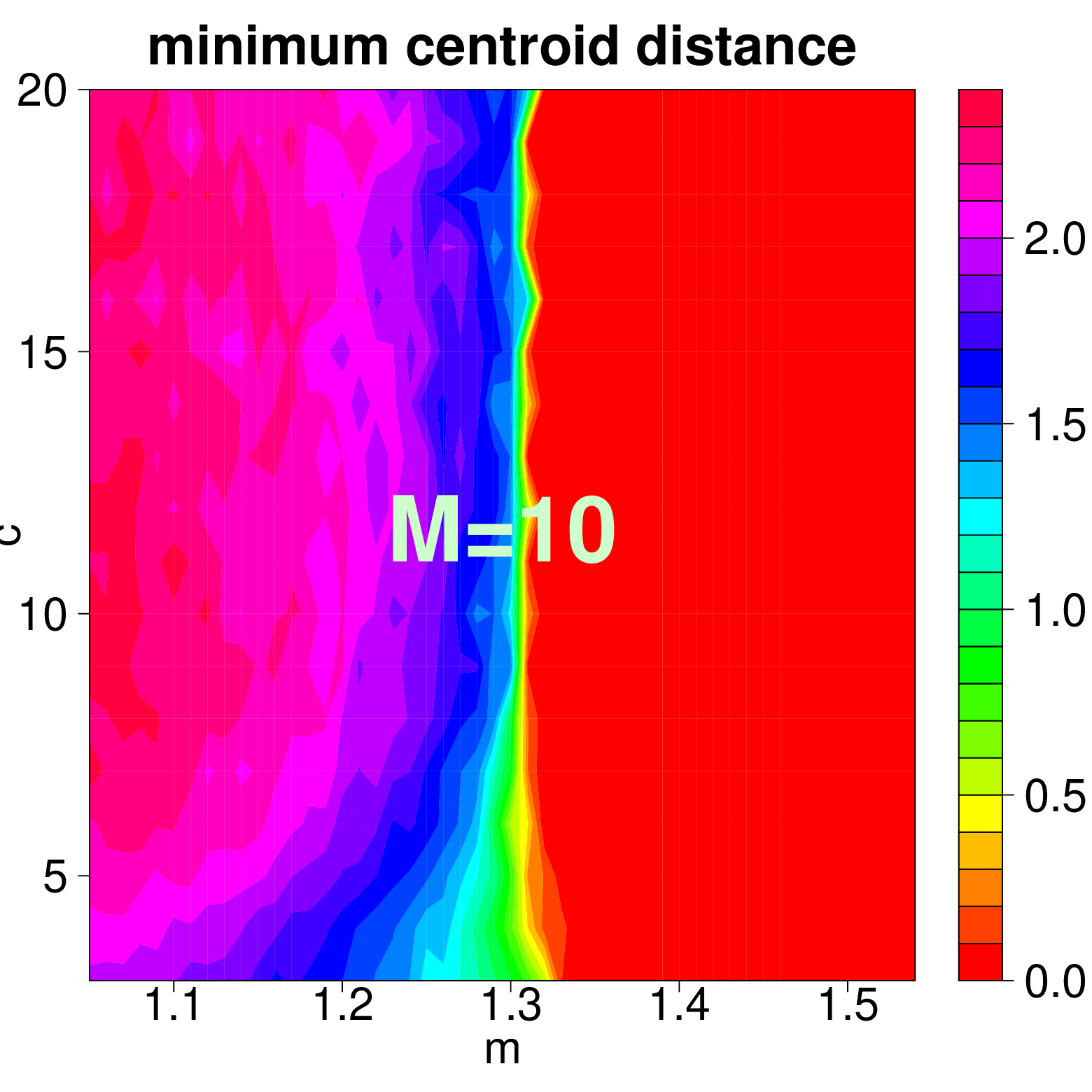}
 \caption{Landscape of the minimum centroid distance for randomized data sets with Gaussian-distributed
clusters. The numbers of previously set different clusters are $M=1$,2,5 and 10. The data sets have the same total
number of objects, $N=1000$,
the dimension of the sets is 10 and we have $w=1$. No significant
differences can be observed except for the panel with $M=1$ where the threshold $m_t$ seems to have a slightly larger
value.}
\label{fig:c5}
\end{figure*}

Fig.~\ref{fig:c5} compares the minimum centroid distance for differently distributed data sets, each randomized before
applying the cluster validation. The picture remains mainly the same, with exception of the case $M=1$, where the
threshold $m_t$ lies at a slightly higher value. The reason is that the threshold still varies within some range for
randomized
data sets of equal dimension and number of objects. The magnitude of this variation decreases for
high-dimensional data sets. 

Despite the normalization of each object to have standard deviation 1 and
mean 0, a
strong bias of the values towards certain dimensions may occur. This bias leads to different results
for the
clustering of the randomized data set. By processing different data sets with the same parameters but different
positions of the artificial Gaussian--distributed objects' center, we try to capture the
effects of both
symmetric as well as biased data sets. The case $M=1$ in Fig.~\ref{fig:c5} corresponds to the clustering results of
mostly strongly biased data. The bias becomes large the more the center of the Gaussian deviates from the origin of the
coordinate system. For $M>1$, this bias becomes smoothed out by randomization and therefore $m_t$
varies much less. For example, a biased data set would be gene expression data where most of the genes are
up-regulated at one of the experimental stages (dimensions).

\begin{figure*}[htb]
 \centering
 \includegraphics[angle=270,width=0.21\textwidth]{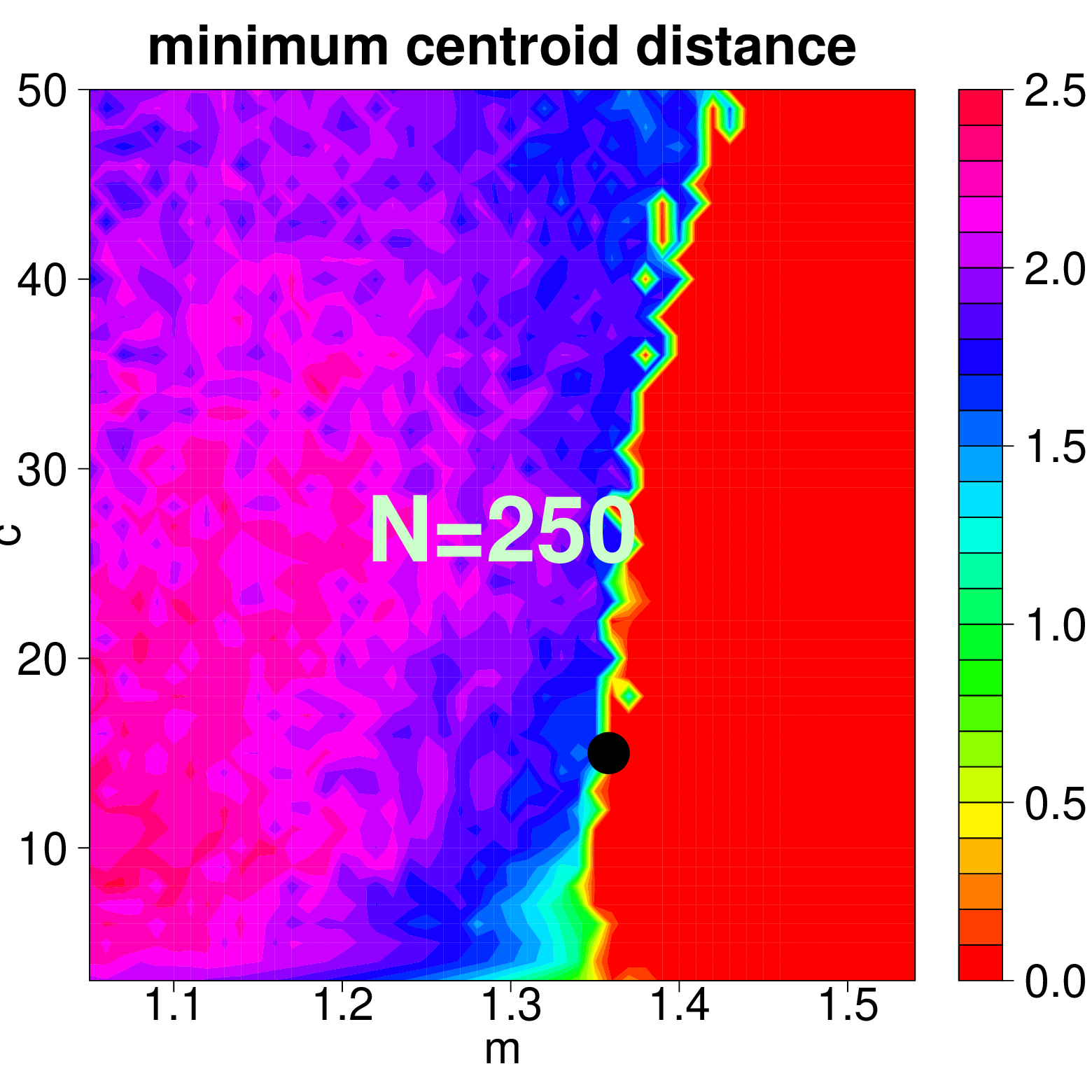}
 \includegraphics[angle=270,width=0.21\textwidth]{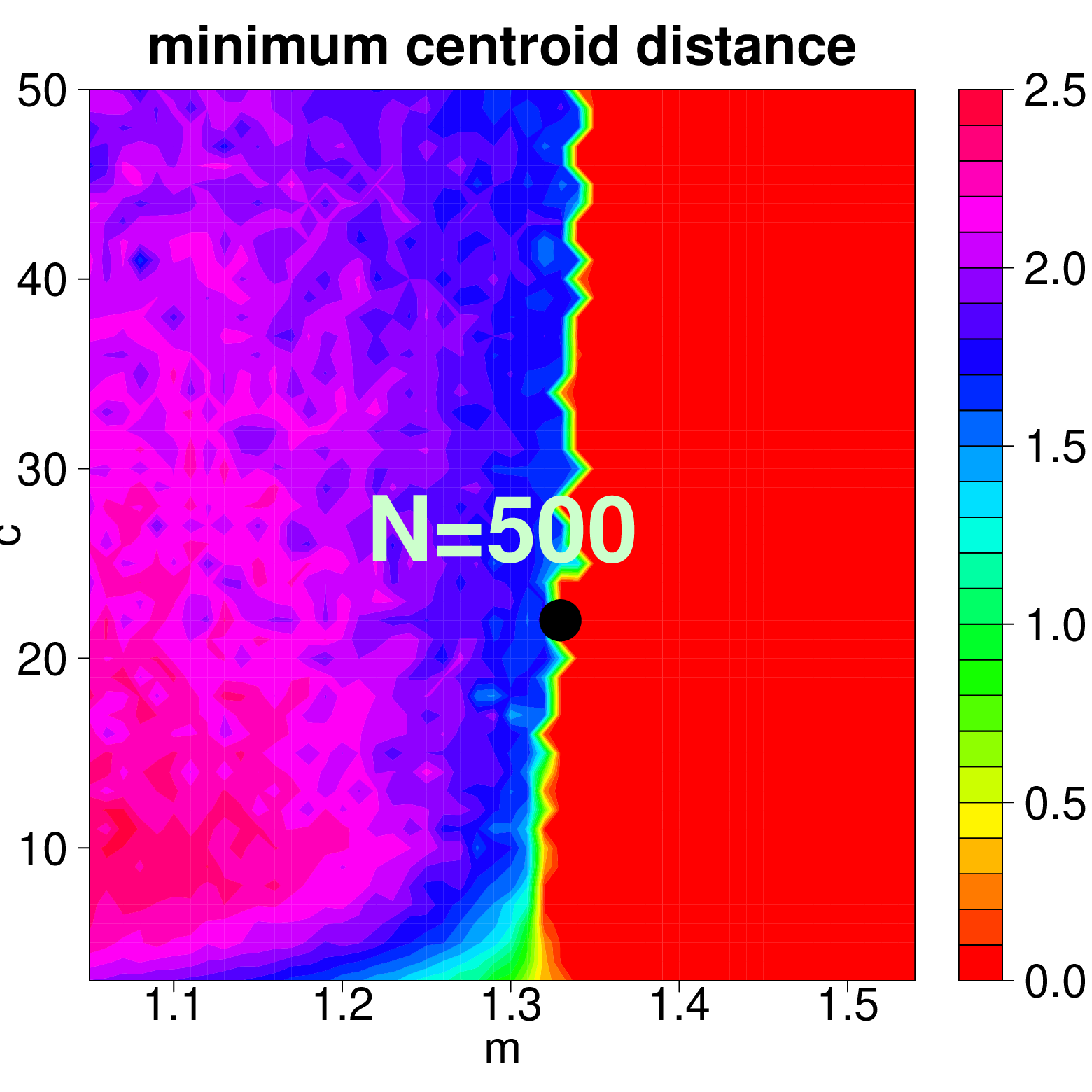}
 \includegraphics[angle=270,width=0.21\textwidth]{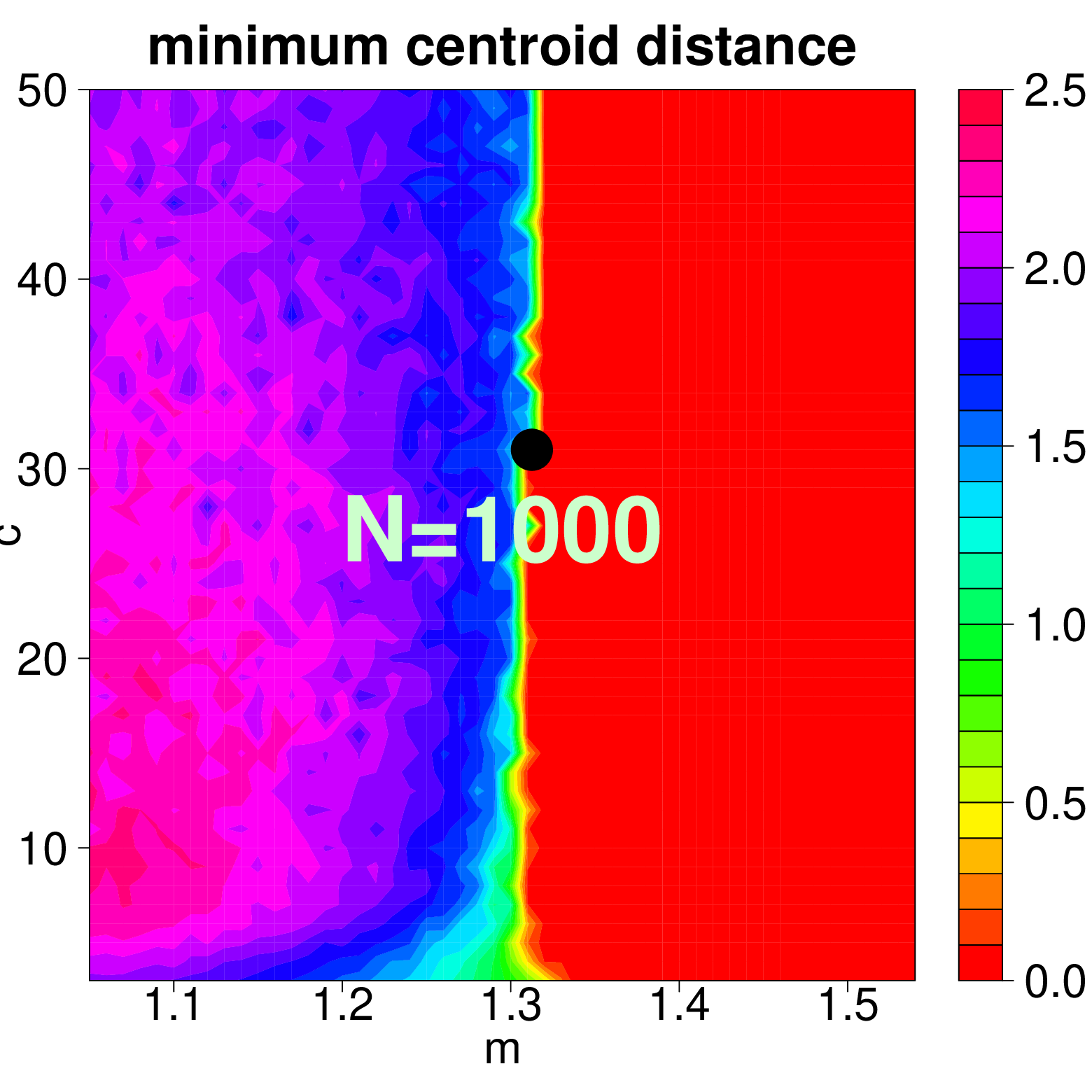}
 \includegraphics[angle=270,width=0.21\textwidth]{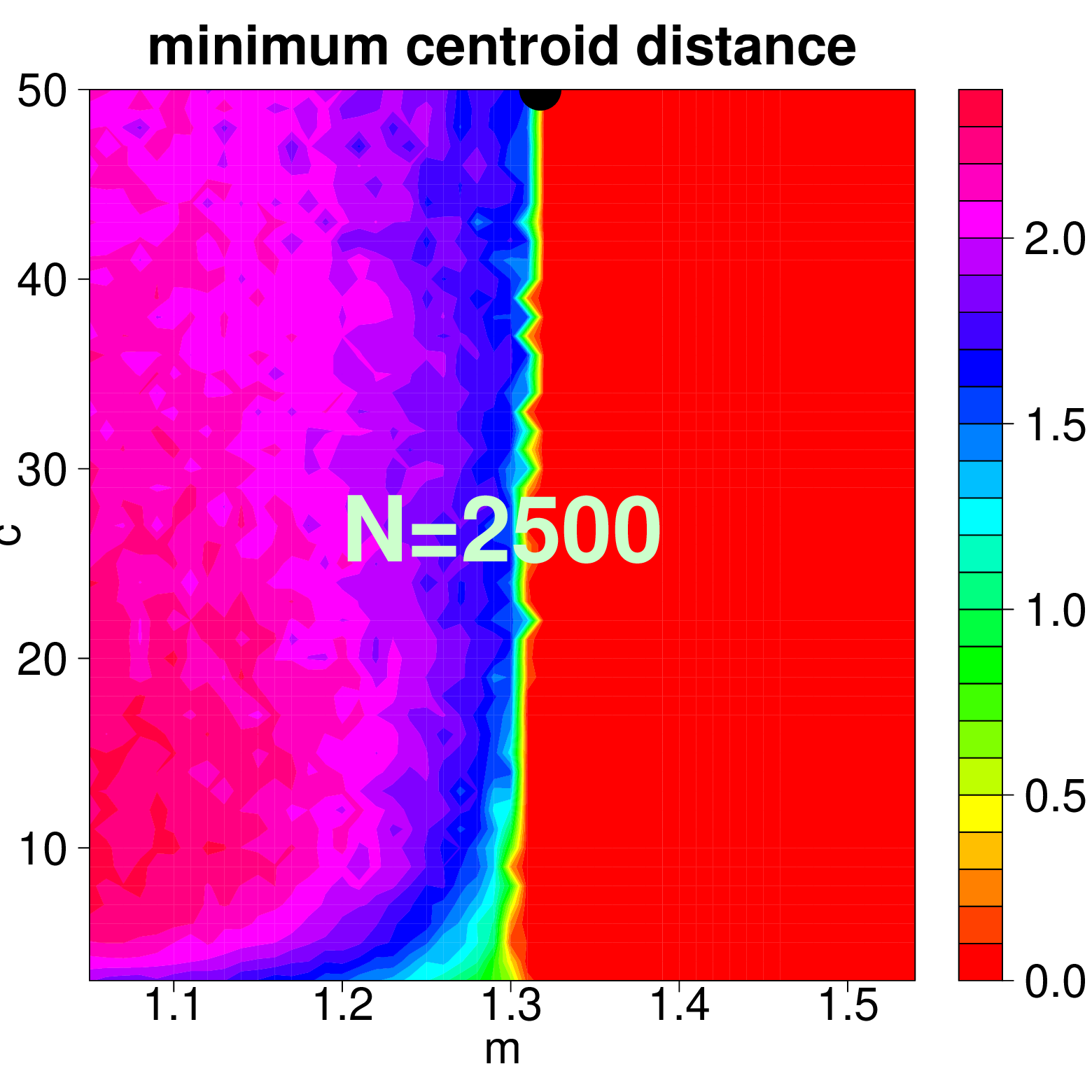}
 \caption{Landscape of the minimum centroid distance from randomized data sets with different numbers of objects,
$N=250,500,1000,2500$. The threshold $m_t$ decreases for increasing $N$ and seems to saturate for very large numbers.
We took $D=10$, $w=1$ and $M=5$. The black points indicate the position where we take the fuzzifier threshold $m_t$.}
\label{fig:c6}
\end{figure*}

The analysis of the simplified model showed also a dependency of the fuzziness in the data set on the number of
objects, $N$, although weaker than the one on the dimension of the data set.
Fig.~\ref{fig:c6} confirms this result, showing that $m_t$ increases for smaller $N$ and saturates at
a certain level for large $N\gtrsim 1000$.

\section{Estimating the optimal value of the fuzzifier}
\label{sec:fuz}
\noindent
We now focus on the estimation of the dependency of the threshold on both $N$ and $D$, i.e. we neglect the
effect of biased data sets. This threshold will then be taken as the optimal value.
A rule of thumb for the maximum number of clusters in a data set is that
it does not exceed the square root of the number of objects~\citep{Zadeh65}.
As the threshold of the minimum of centroid distances $m_t$ does not vary with $c$, we determine the threshold in the
following analysis by carrying out cluster validations with different $m$ for $c=\sqrt{N}$. Precisely, the threshold
$m_{t}$ corresponds to the value of the fuzzifier at which
the
minimum centroid distance falls below $0.1$ for the first time. Note, that we
hereby exclude the situation that the centroids of two clusters locate at
mutually small distances of less than $0.1$. However, this limitation did not affect the results.

The clustering is carried out 5-10 times, each validation for a
different randomized artificial data set having the same parameters. From these different
runs we take
the largest value of $m_t$.

The usage of $m=m_t$ in the cluster validation of the original data set has two advantages. First, a data set lacking
non-random clusters does not provide any reasonable results, i.e. the number of detected hard clusters is lower than
the parameter $c$. This means that the value of the minimum centroid distance is around zero for all $c$. Second, this
smallest
allowed value of $m$ guarantees an optimal estimation of a maximal number of clusters which is in general better than
for larger $m$ and so still ensures the recognition of barely detectable clusters. 

\begin{figure}[htb]
 \centering
 \includegraphics[angle=270,width=0.35\textwidth]{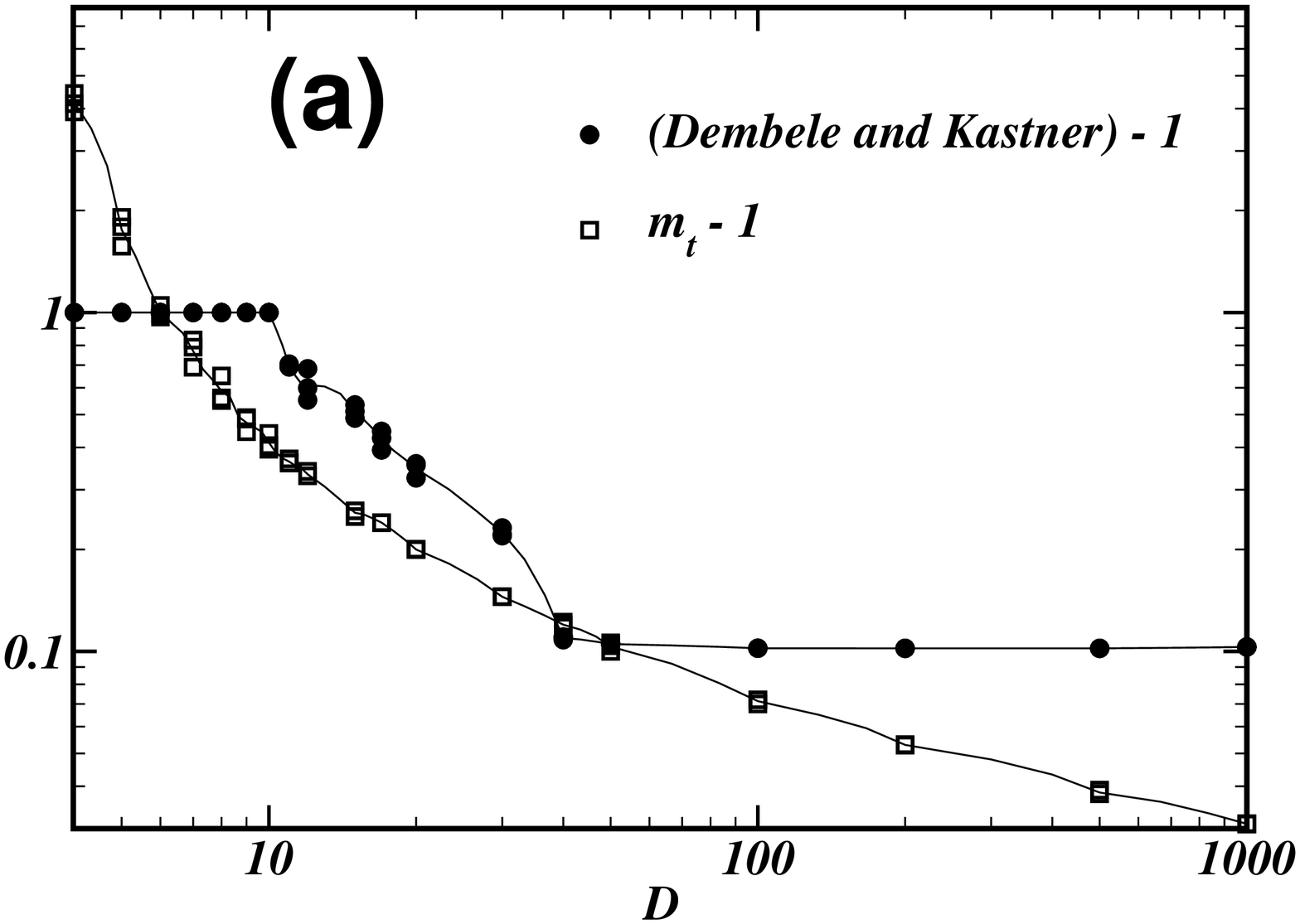}
 \includegraphics[angle=270,width=0.35\textwidth]{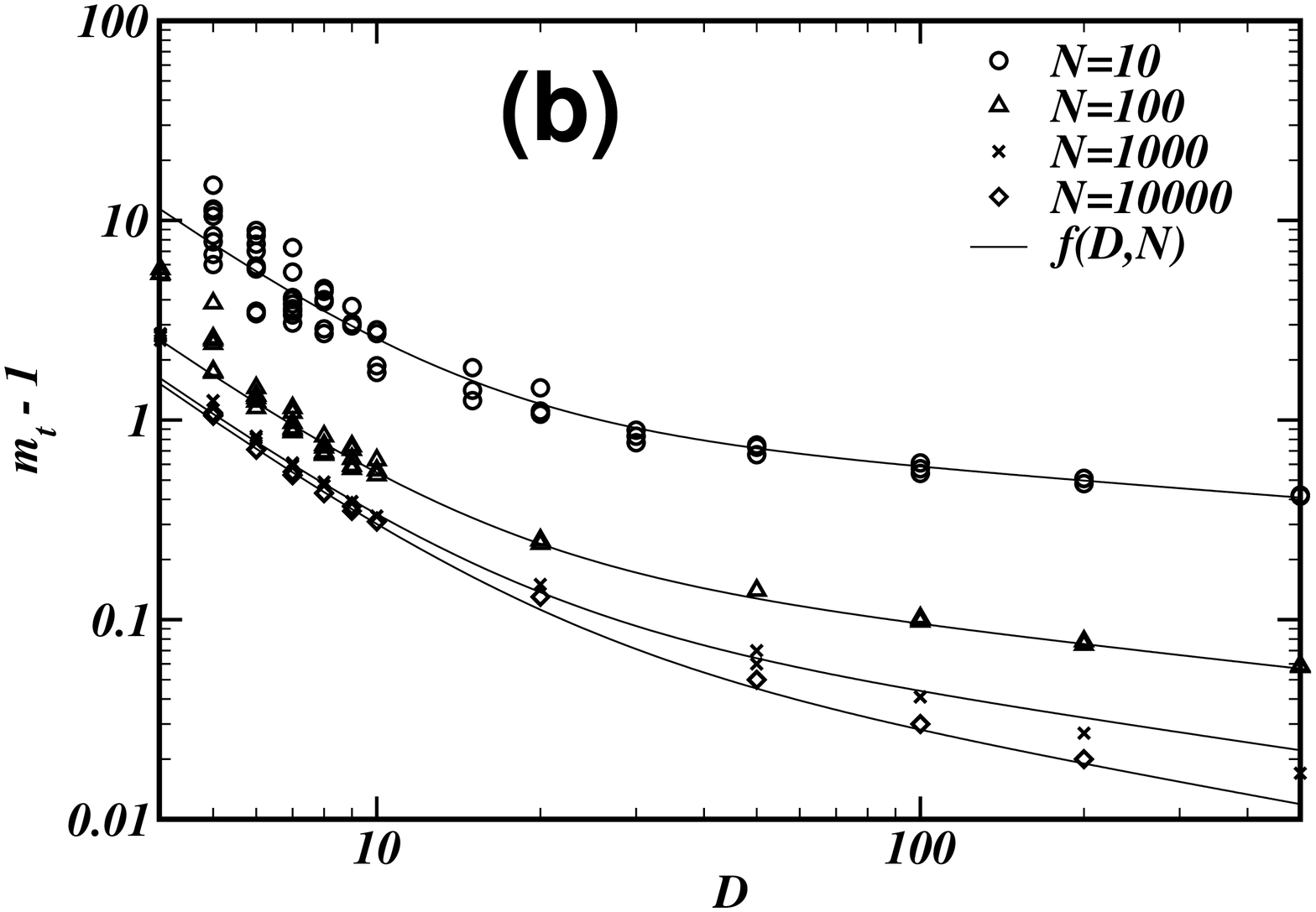}
\caption{{\bf (a)}: A comparison of our method for the estimation of the fuzzifier to the one presented in
\citet{Dembele2003} shows that
the value of $m$ is mostly overestimated in the latter. In addition, our method allows to cope with a larger
dimensional range. {\bf (b)}: Comparing the threshold of the
minimal centroid distance for randomized data sets with different numbers of
objects, $N$. The threshold increases for larger $N$ and the curve seems to approach a limiting shape for very large
$N$. Fluctuations become large for $D<10$. The lines show the values of the fitting function, Eq.~(\ref{eq:emp}).}
\label{fig:mt1}
\end{figure}

The dependency of $m_t$ on the dimension of the data set is shown in Fig.~\ref{fig:mt1}a and compared to the
values calculated by the method introduced in \citet{Dembele2003}. The curves from the latter method exhibit the
same tendency but an overestimation of the fuzzifier. 

A thorough analysis, calculating $m_t$ for randomized data sets of different dimensions and object numbers shows
a general functional relation between $m_t$ and the data set properties. 
The following function provides a good fit of the curves for all combinations of $N$ and $D$,
\begin{eqnarray}
  \label{eq:emp}
  f(D,N) &=& 1+ \left(\frac{1418}{N}+22.05\right)D^{-2}\nonumber\\
&&+\left(\frac{12.33}{N}+0.243\right)D^{-0.0406\ln(N)-0.1134}~.
\end{eqnarray}
Both the data points of $m_t$ and their empirical fit with Eq.~(\ref{eq:emp}) are depicted in Fig.~\ref{fig:mt1}b. The
prediction with the empirical formula improves for large $N$ and large $D$. For smaller values of these input
values, the $m_t$ obtained from the artificial sets may deviate from the predicted value due to their dependency on
the individual data set. 

\begin{figure}[htb]
 \centering
 \includegraphics[angle=270,width=0.3\textwidth]{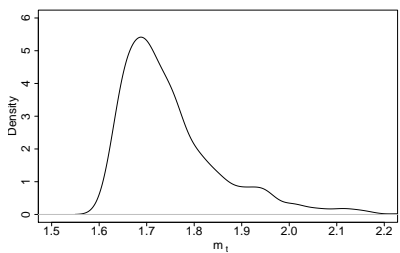}
\caption{The density distribution of threshold values for different implementations of randomized artificial data sets
with $M=1$,
$D=7$, $w=0.1$ and $N=200$. }
\label{fig:mt3}
\end{figure}

We calculated the density distribution of $m_t$ for artificial sets with the same parameters, setting
$M=1$, $D=7$ and $N=200$ (Fig.~\ref{fig:mt3}). The corresponding prediction for $m_t$ is given by
$f(7,200) = 1.75$. The only difference between the data sets consists in the position of the mean of the
Gaussian, and thus the bias of the data. The maximum of the distribution lies
at a slightly smaller value than the one predicted in Eq.~(\ref{eq:emp}). The figure shows also that the lower limit of
$m_t$ is rather well defined whereas high
values are possible, even far away from the maximum. Consequently, for data sets with small $N$ and
$D$, Eq.~(\ref{eq:emp}) may be more useful for the
estimation of the lower limit of $m_t$ than for its exact prediction. However, the prediction works
much better for larger values of $D$ and $N$ where computational time becomes an issue.

\begin{table}[htb]
\processtable{Comparing estimated values of $m_t$ to their predictions from Eq.~(\ref{eq:emp}).\label{table:2}}
{\begin{tabular}{cccccc}\toprule
 Data set & ~$D$~ & ~$N$~ & ~$m_t$~ & $f(D,N)$\\\midrule
iTRAQ1, suppl. table 1~\citep{Pierce2008}& 7 & ~1886~ & ~1.54~ & 1.56 \\
iTRAQ2, suppl. table 2~\citep{Pierce2008}  & 7 & 829 & 1.56 & 1.59 \\
iTRAQ3, Table 4~\citep{Wolf-Yadlin2007} & 7 & 222 & 1.81 & 1.73  \\
Ecoli~\citep{Horton96} & 7 & 336 & 1.64 & 1.67  \\
Abalone~\citep{Nash94} & 8 & 4177 & 1.41 & 1.44\\
Serum~\citep{Iyer99} & 13 & 517 & 1.27 & 1.25\\
Yeast1~\citep{Tavazoie99} & 16 & 2885 & 1.18 & 1.16\\
Yeast2~\citep{Cho98} & 17 & 2951 & 1.17 & 1.15\\
Ionosphere~\citep{Sigillito89} & ~34~ & 351 & 1.13 & 1.1\\\botrule
\end{tabular}}{}
\end{table}

Eq.~(\ref{eq:emp}) accounts also for randomized real data sets where the distribution within a cluster may be non-Gaussian. For the analysis, we tested data sets from different origin
including biological data from protein research~\citep{Horton96,Pierce2008,Wolf-Yadlin2007}, microarray
data~\citep{Iyer99,Tavazoie99,Cho98} and data gathered from non-biological
experiments~\citep{Nash94,Sigillito89}. 

Table~\ref{table:2} compares the minimum centroid threshold calculated from
the randomized data sets to the empirical value obtained from Eq.~(\ref{eq:emp}). We find a deviation for the iTRAQ3
data set
having a small $D=7$ and $N=222$. From Fig.~\ref{fig:mt3} we see that the higher value of
$m_t=1.81$ is still within the range of the distribution. Note, that the optimal fuzzifier value for the yeast2 data
set was estimated to be $m=1.15$ in \citet{Futschik2005}, identical with our estimation. 

\section{Determining the number of clusters}
\label{sec:clu}
\begin{figure}[htb]
 \centering
 \includegraphics[angle=270,width=0.15\textwidth]{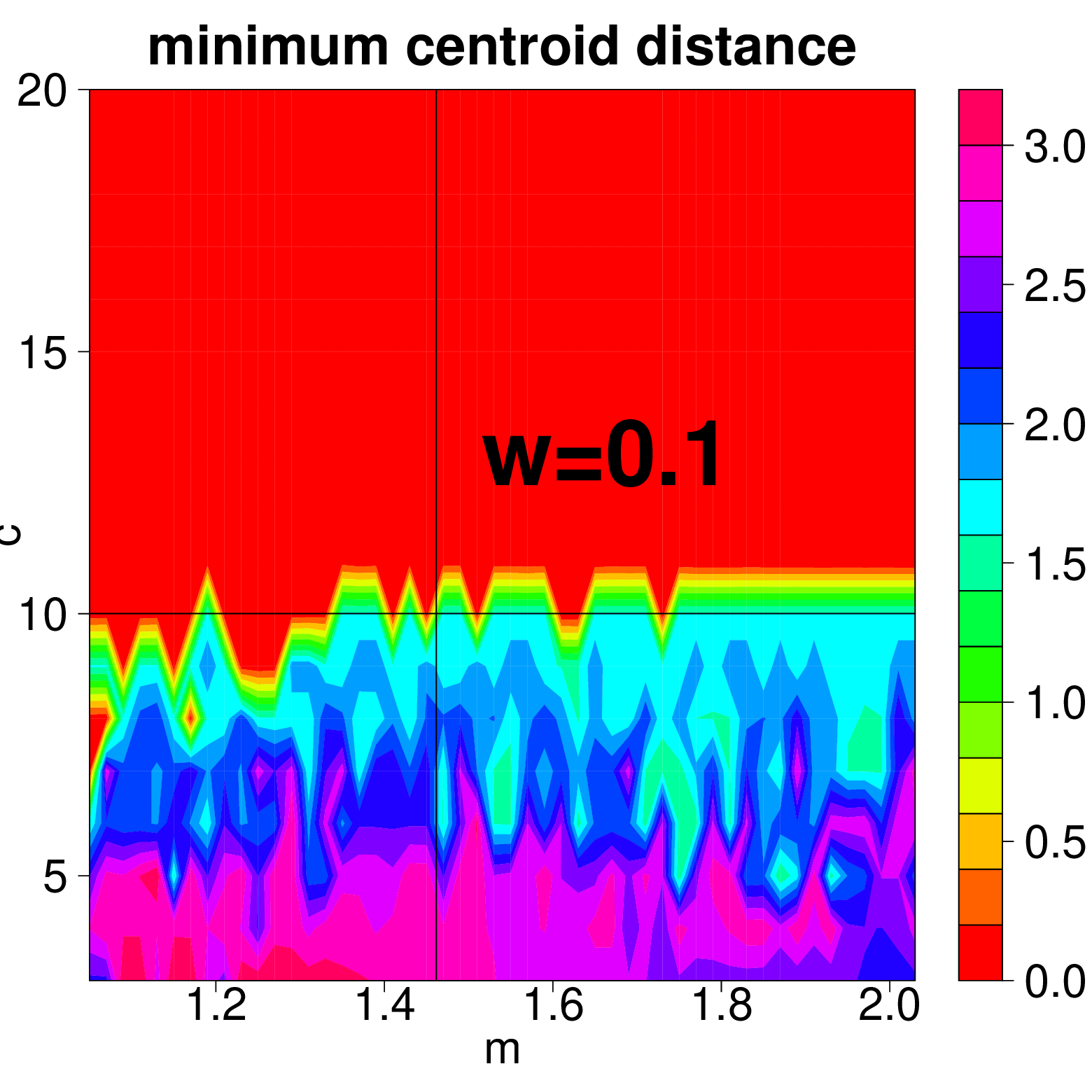}
 \includegraphics[angle=270,width=0.15\textwidth]{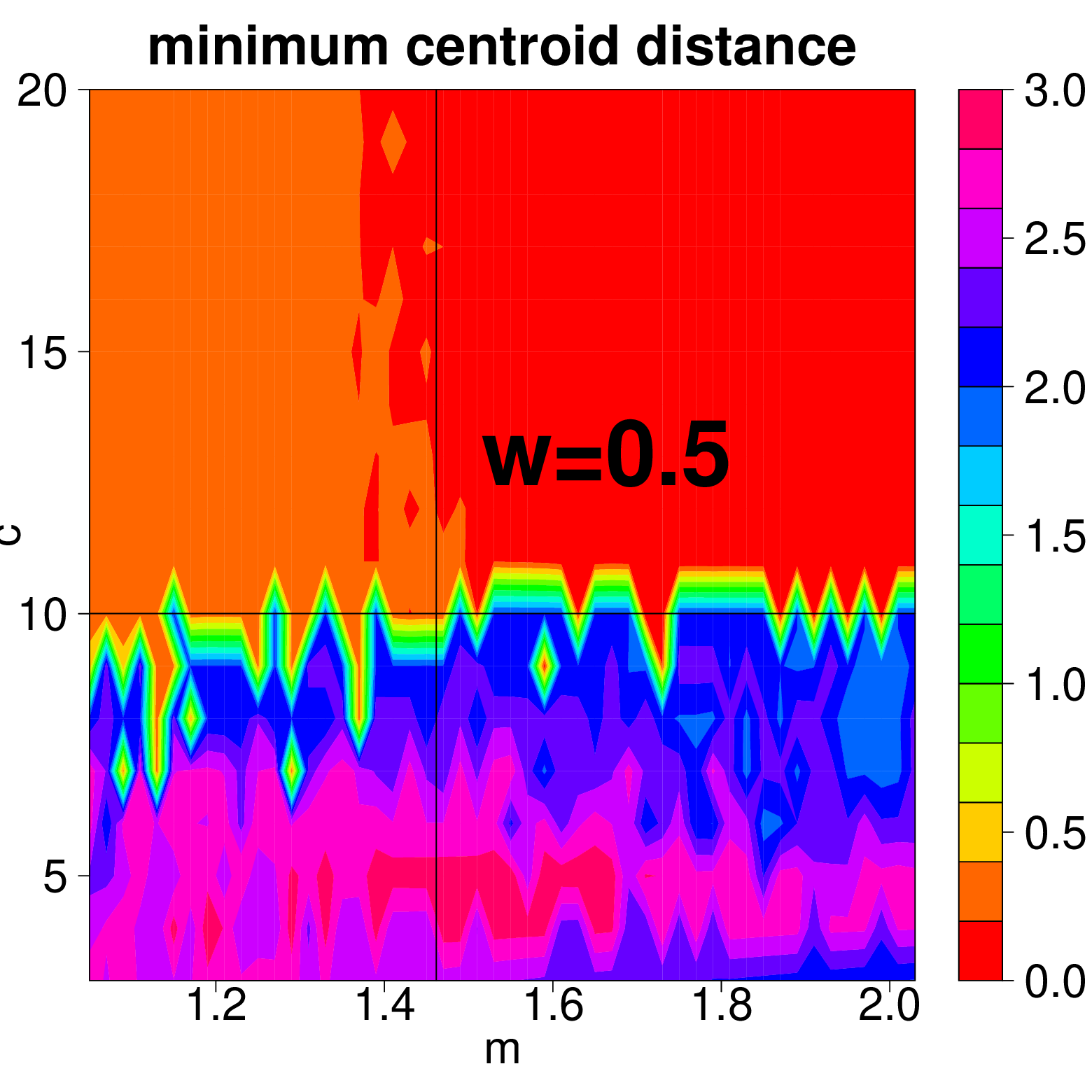}
 \includegraphics[angle=270,width=0.15\textwidth]{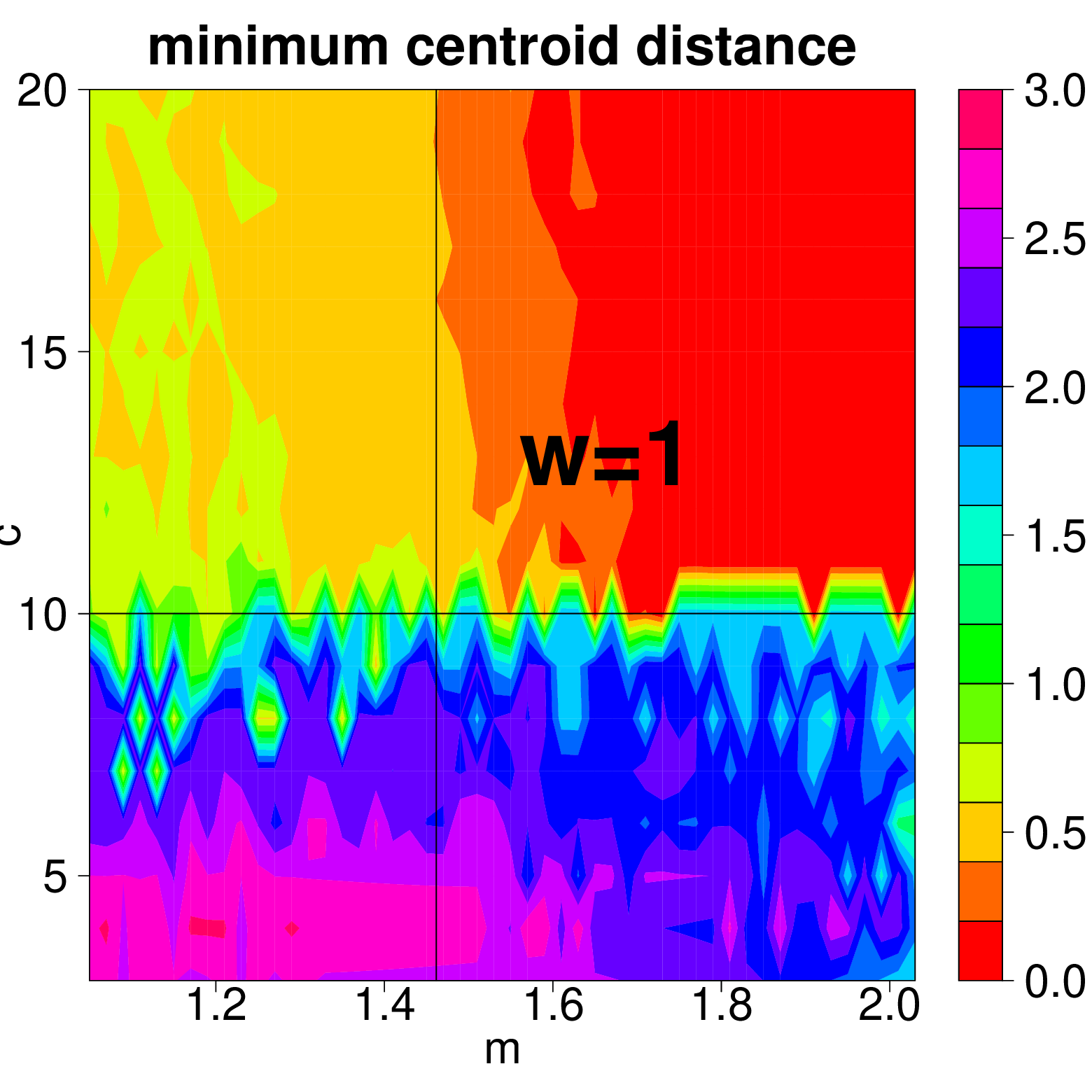}
 \includegraphics[angle=270,width=0.15\textwidth]{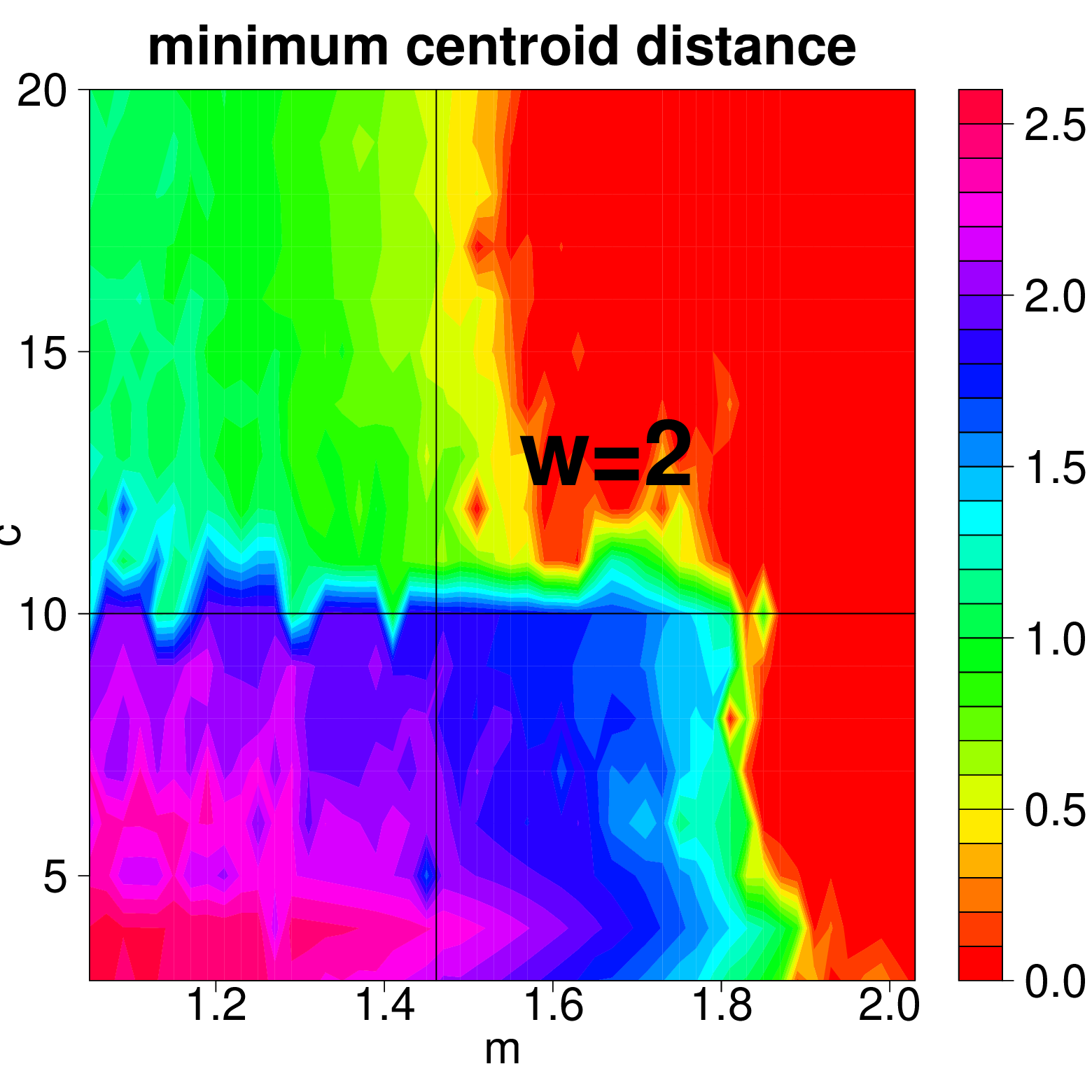}
 \includegraphics[angle=270,width=0.15\textwidth]{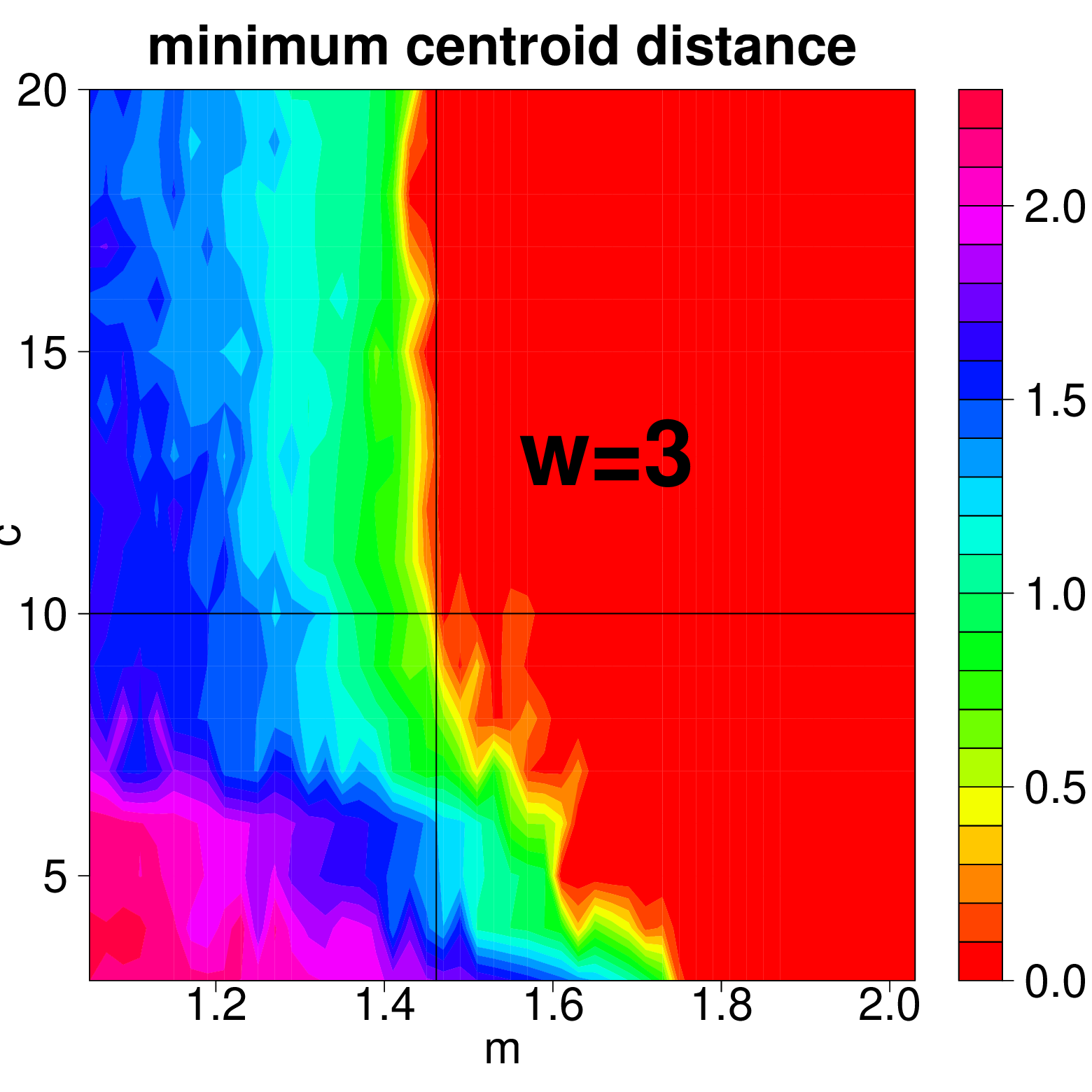}
 \includegraphics[angle=270,width=0.15\textwidth]{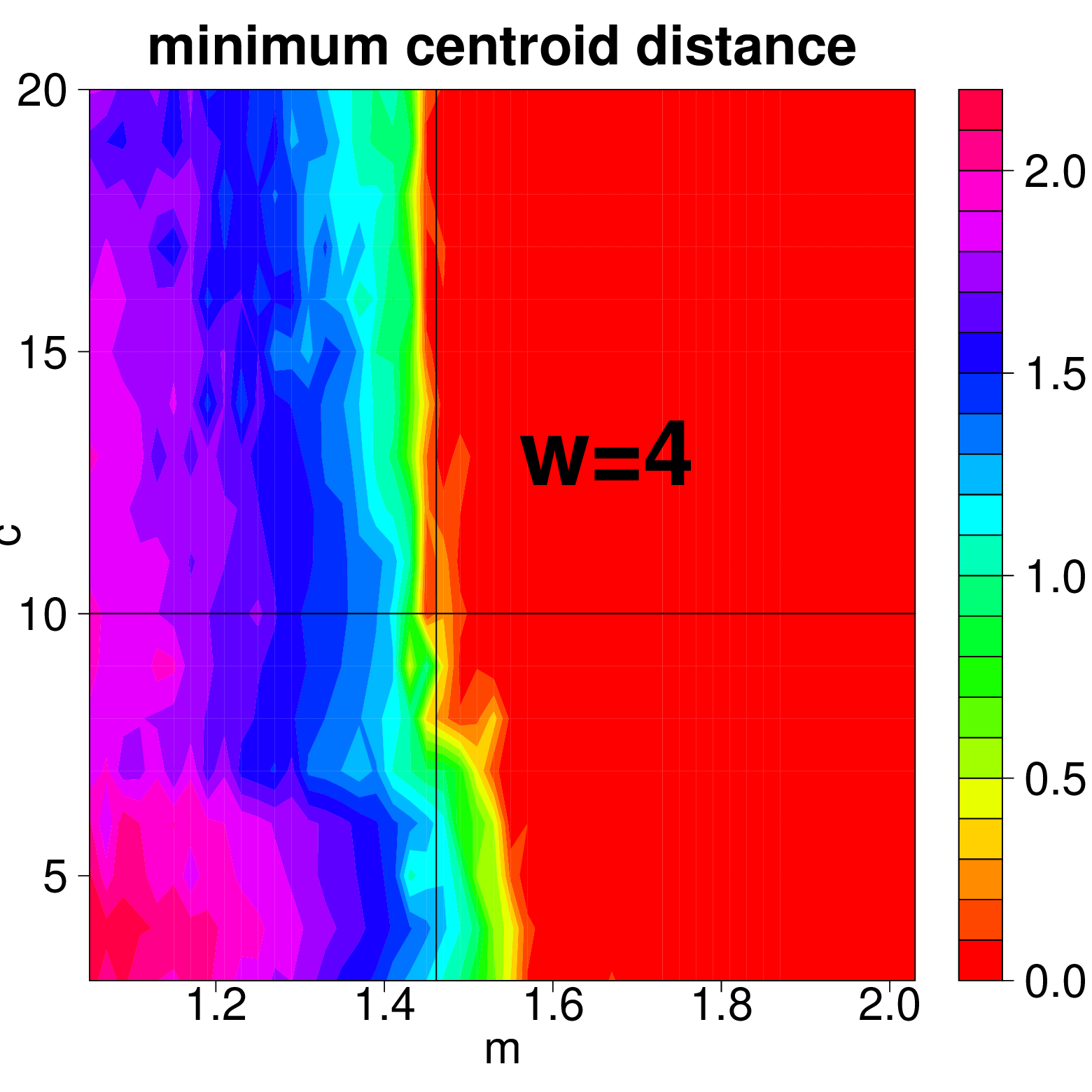}
\caption{Landscape of the minimum centroid distances for a set of 10 clusters with each having 100 objects of $D=8$.
The clusters were produced to have a Gaussian distribution with different standard deviations, being $w=0.1,0.5,1$
(upper
panels left, middle and right), and $w=2,3,4$ (lower panels left, middle and right). The black lines indicate $m_t$ and
$c_t$.}
\label{fig:d1}
\end{figure}
\noindent
After calculating the optimal value of the fuzzifier by either using Eq.~(\ref{eq:emp}) or determining $m_t$ directly as
done above, the final step consists in estimating the number of clusters in the data set. Various validity indices
for the quality of the clustering are present in the literature. They in general are a function of the
membership values, the centroid coordinates and the data set. The results for the indices summarized in
Table~\ref{table:indices} will be compared for artificial and real data sets.
\begin{table*}[b]
\processtable{Summary of the validation indices.\label{table:indices}}
{\begin{tabular}{cc}\toprule
Partition coefficient~\citep{Bezdek75} & $V_{PC} = \frac{1}{N}\sum \limits_{i=1}^c \sum \limits_{j=1}^N (u_{ij})^2$ \\
Modified partition coefficient~\citep{Dave96} & $V_{MPC} = 1 - \frac{c}{c-1} (1-V_{PC})$\\
Partition entropy~\citep{Bezdek74} & $V_{PE} =  -\frac{1}{N}\sum \limits_{i=1}^c \sum \limits_{j=1}^N
u_{ij}\log(u_{ij})$  \\
Av. within--cluster distance~\citep{Krishnapuram92} & $V_{\text{AVCD}} = \frac{1}{c}  \frac{1}{N}\sum \limits_{i=1}^c
\frac{\sum
\limits_{j=1}^N  (u_{ij})^m |\mathbf{x_j}-\mathbf{c_i}|^2 }{\sum \limits_{j=1}^N  (u_{ij})^m} $\\
Fukuyama-Sugeno index~\citep{Fukuyama89} & $V_{FS} =\sum \limits_{i=1}^c \sum \limits_{j=1}^N (u_{ij})^m
\left( \left|\mathbf{x}_j-\mathbf{c}_i\right|^2 - \left|\left( \frac{1}{N} \sum \limits_{k=1}^N \mathbf{x}_k
\right)-\mathbf{c}_i\right|^2 \right)$\\
Xie-Beni index~\citep{Xie91} & $V_{XB} = \frac{\sum \limits_{i=1}^c \sum \limits_{j=1}^N \left(u_{ij}\right)^m
\left|\mathbf{x}_j-\mathbf{c}_i\right|^2}{N \min\limits_{i\neq j} \left| \mathbf{c}_i -\mathbf{c}_j \right|^2}$\\
PCAES~\citep{Wu2005} & $V_{\text{PCAES}} =\sum \limits_{i=1}^c \sum \limits_{j=1}^N
\frac{(u_{ij})^2}{\min\limits_{1\leq i\leq c}\left(\sum \limits_{k=1}^N (u_{ij})^2 \right)} - \sum \limits_{i=1}^c
\exp \left( -\min\limits_{k\neq i} \left( \frac{c |\mathbf{x}_i-\mathbf{x}_k|^2}{\sum \limits_{l=1}^c
\left|\mathbf{x}_l-\left( \sum \limits_{s=1}^N \mathbf{x}_s / N \right)\right|^2} \right) \right) $\\
Minimum centroid distance & $V_{MCD} = \min\limits_{i\neq j} \left| \mathbf{c}_i-\mathbf{c}_j \right|^2$\\\botrule
\end{tabular}}{}
\end{table*}

First we take another look on the minimum centroid distance, $V_\text{MCD}$, now taken from the cluster validation
of artificial (not randomized) data sets (Fig.~\ref{fig:d1}). The panels show $V_\text{MCD}$ for data sets with
10 Gaussian--distributed clusters, each panel for a set of Gaussians with different standard deviations. 
For data sets with clearly separated clusters (small standard deviations), the picture is completely different to the
one of a randomized data set (Figs.~\ref{fig:c3}--\ref{fig:c6}). A strong decay,
this time
not necessarily to zero, occurs at $c=c_t$ independent of the value of the used
fuzzifier $m$. Note that in the randomized case the decay was at $m_t$ for all $c$. The position of the
sudden decrease coincides with the number of clusters $M=10$ of the artificial data set, and thus the
minimum centroid distance provides a reasonable measure also to determine the optimal number of clusters.  For
more mixed clusters, the landscape transforms gradually into the picture observed for randomized sets.

The parameter landscapes of real data sets will exhibit a combination of two extremes, a plateau below the threshold
$c_t$ for a data set with clearly distinguishable cluster and a plateau below $m_t$ for a completely noisy data set.
We can also observe that  the number of found clusters decreases with increasing $m$ (cases $w=2$, $w=3$ and $w=4$ in
Fig.~\ref{fig:d1}) as would be expected.
\begin{figure}[!htb]
 \centering
 \includegraphics[angle=0,width=0.5\textwidth]{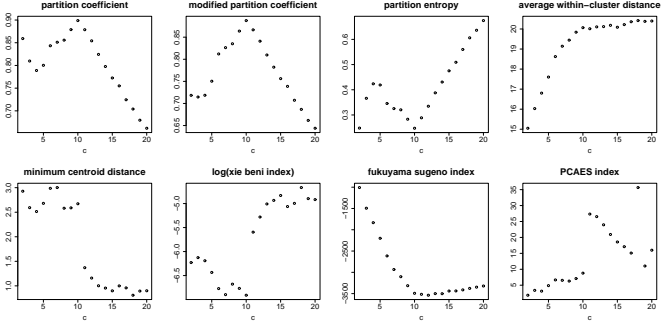}
\caption{Comparison of the different validity indices for an artificial system of 500 10-dimensional data points, with
10 clusters
each with 50 points. All indices show that $c=10$ is the optimal solution.}
\label{fig:d2}
\end{figure}

Eq.~(\ref{eq:emp}) gives $m_t= 1.47$ for the parameters of the artificial data sets in Fig.~\ref{fig:d1}. The
figure shows that some of the clusters may be recognized even for $w=3$ and $w=4$, when using $m=m_t$ for the
clustering (i.e. we find a decay of the minimum centroid distance at $c=7$ for $m=1.47$). For larger values of
the fuzzifier, no clusters can be
detected, whereas the decay begins to become less accentuated for smaller
$m$-values. Hence, the minimum centroid distances may be considered as a
powerful validity index for the case that the appropriate $m=m_t$ is chosen. Another
advantage of using $V_\text{MCD}$ is that its calculation is faster
than the one of the other validity indices.

\begin{figure}[!htb]
 \centering
 \includegraphics[angle=0,width=0.5\textwidth]{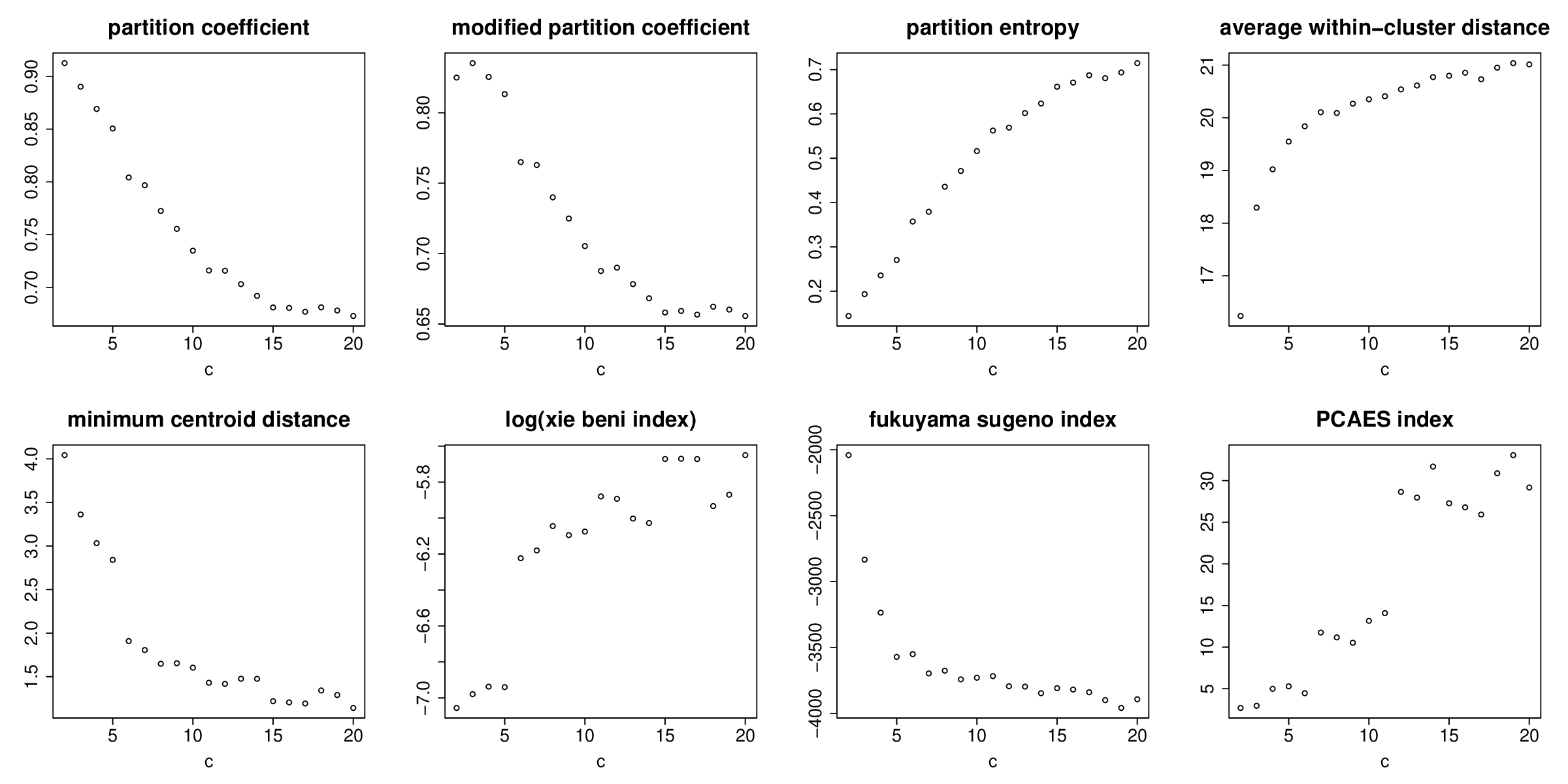}
\caption{Comparison of the different validity indices for the serum data set. For real data, it is obviously
more difficult to estimate the number of clusters. However, some indices have a jump at or near $c=5$. }
\label{fig:d3}
\end{figure}

For a comparison of the different validation indices, we generated a data set with $D=13$, $N=500$, $M=10$ and $w=2$
for which Eq.~(\ref{eq:emp}) gives $m_t=1.25$. Fig.~\ref{fig:d2} shows the validation indices versus the
cluster number $c$ using $m=m_t$. All methods clearly indicate $c=10$ as the optimal solution. Note, that there is also
a strong
decay of $V_{\text{MCD}}$ at $c=10$.

\begin{figure}[htb]
 \centering
 \includegraphics[angle=270,width=0.2\textwidth]{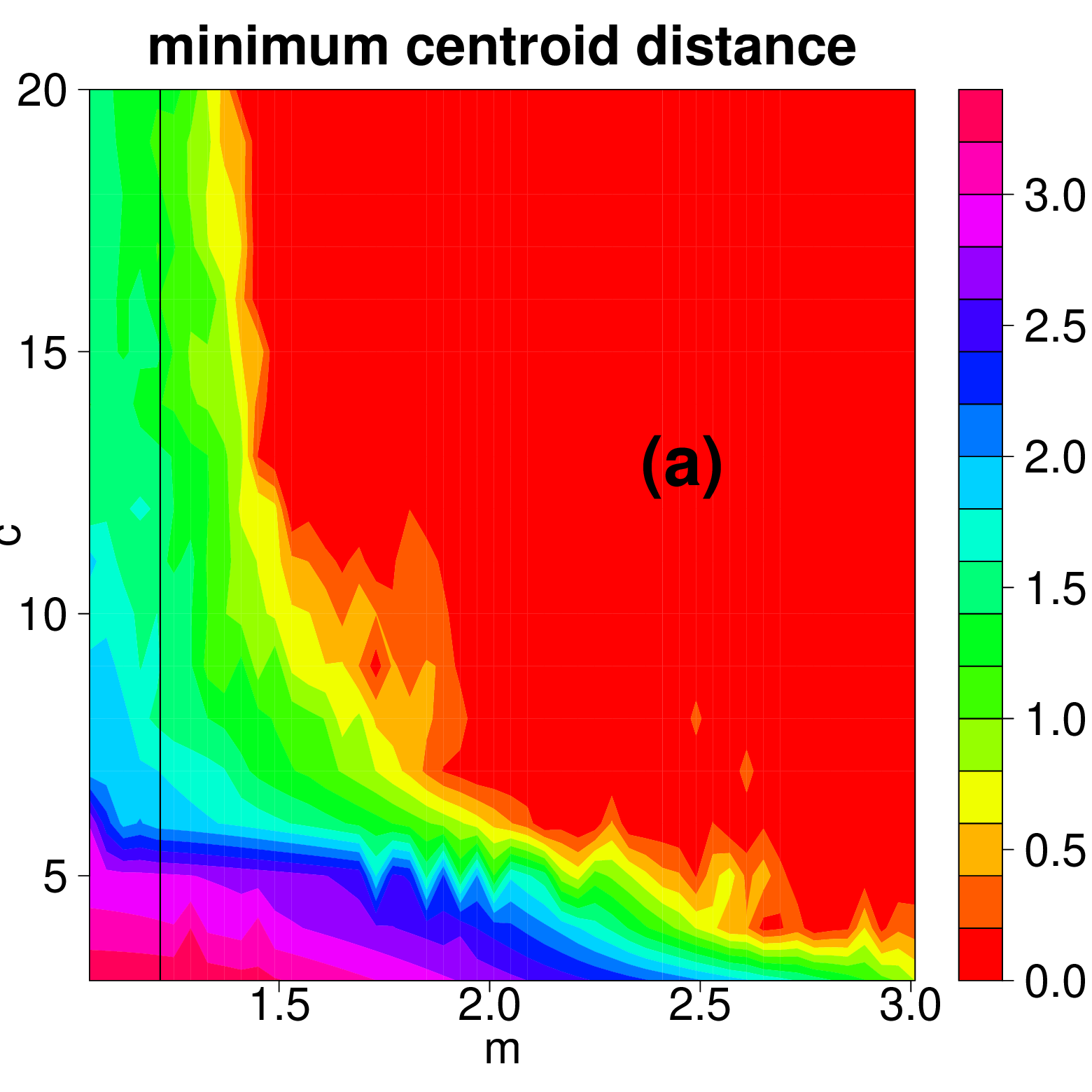}
 \includegraphics[angle=270,width=0.45\textwidth]{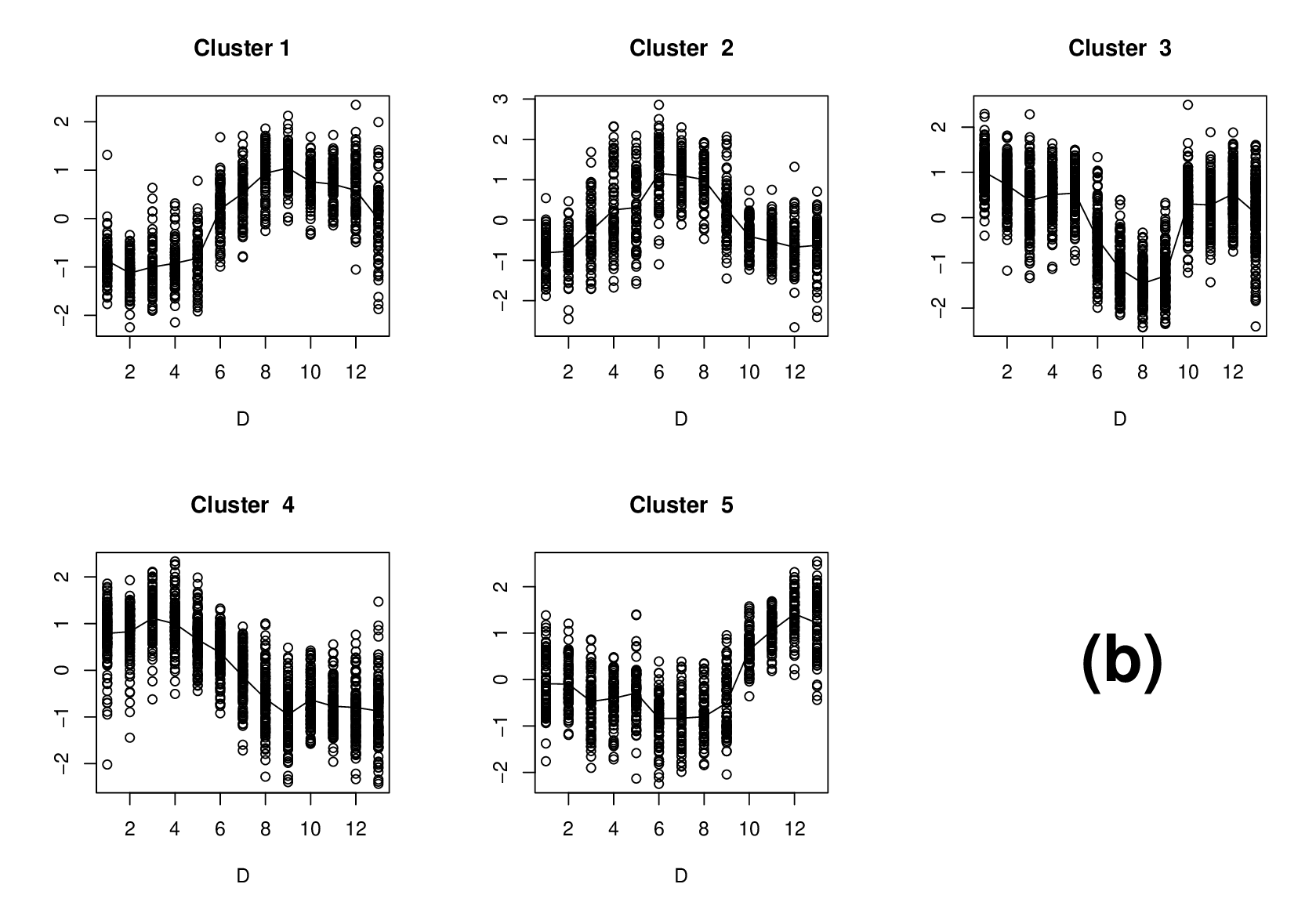}
\caption{{\bf (a)}: Landscape of the minimum centroid distances for the serum data set. The strongest decay is found for
$c=5$ around $m_t=1.25$. The black line denotes $m_t=1.25$, {\bf (b)}: Patterns of the objects in all 5 clusters
depicting only the ones with membership values larger
than $1/2$. The centroids are shown by the lines.}
\label{fig:d4}
\end{figure}

Real data normally is more complex than the artificial sets analyzed here. Not only the kind of noise may be different
but also the clusters may not have normal-distributed values and the
clusters might have different sizes. As a
consequence, often an optimal parameter set does not exist, and the most appropriate solution must
be chosen manually out of the best candidates. 
As a test data set we used the serum set~\citep{Iyer99} that has the same number of dimensions and a similar number of
objects as the
artificial data set analyzed in Fig.~\ref{fig:d2}. The validation indices now do not
agree in giving a clear indication for the number of clusters
in the system (Fig.~\ref{fig:d3}). However, most of them yield $c=5$ as the optimal solution. The
abrupt decay of the minimum centroid distance at the same $c=5$ is remarkable. Fig.~\ref{fig:d4}a depicts the  landscape of the
minimum centroid distance for the serum data set over a large range of $m$ and $c$. First, we observe a similarity
between Fig.~\ref{fig:d4}a and the case $w=3$ in Fig.~\ref{fig:d1} suggesting that the data set consists of
overlapping but distinguishable clusters. The minimum centroid distance has a plateau for $c\leq 5$ and $m<2$ with a
decay at $c=5$ over a considerable range of $m$-values around $m_t=1.25$ indicating $c=5$ as the optimal choice.

Fig.~\ref{fig:d4}b shows the patterns of all clusters for the cluster validation on the serum data set taking
$c=5$ and $m=1.25$. The lines correspond to the coordinates of the centroids. Only objects with membership values over
$1/2$  for the corresponding cluster are shown.
%

\section{Conclusions}

In fuzzy c-means cluster validation, it is crucial to choose the optimal parameters since a large fuzzifier value leads to
loss
of information and a low one leads to the inclusion of false observations originating from random noise. The value
of the fuzzifier was frequently set to 2 in many studies without specification of the amount of noise in the system. We
show here that the strong dependence of the optimal fuzzifier value on the dimension of the system requires
fine--tuning of this parameter. 

To our knowledge, two methods exist to obtain the fuzzifier by processing the data
set~\citep{Dembele2003,Futschik2005}. 
We present here a new, fast and simple method to estimate the fuzzifier being calculated from only two main properties
of the data set, its dimension and the number of objects. Using this method, we obtained not only an
optimal balance between maximal cluster detection and maximal suppression of random effects but it also allows us to process larger
data sets. 
The results suggest that biased data leads to an increase of the value of
the fuzzifier in low-dimensional data sets with a small number of objects (for instance $N<200$ and $D<8$) and thus the
parameters should be chosen
carefully for this type of data. 
The estimation is based on the evaluation of the minimal distance between the centroids of the clusters found by the
cluster validation.
The minimum centroid distance provides sufficient information for the estimation of the other parameter necessary
for the clustering procedure, the number of clusters, and eliminates the need for calculation of computationally intensive
validation indices. 

In data from proteomic studies, especially labeled mass spectrometry data, protein expressions are compared over
a generally smaller number of stages (for instance less or equal to 8 in iTRAQ data). As our study shows, the optimal
value of the fuzzifier increases strongly at low dimensions to values larger than $m=2$ making it difficult to obtain
well-defined clusters. Therefore, a compromise needs to be made, by allowing lower fuzzifier values, $m<m_t$, 
admitting the influence of random fluctuations to the results. A quantification of the confidence of the cluster
validation of low-dimensional data needs to be carried out or other methods of data comparison, such as direct
comparison of the absolute data values, must complement the data analysis.

\section*{Acknowledgement}

\paragraph{Funding\textcolon} VS was supported by the Danish Council for Independent Research, Natural Sciences (FNU).

\bibliographystyle{natbib}
\bibliography{clusters}

\begin{thebibliography}{}

\bibitem[Babuska(1998)Babuska]{Babuska98}
Babuska, R. (1998).
\newblock {\em Fuzzy Modeling for Control\/}.
\newblock Kluwer Academic Publishers, Dordrecht.

\bibitem[Bezdek(1974)Bezdek]{Bezdek74}
Bezdek, J.~C. (1974).
\newblock {{C}luster validity with fuzzy sets}.
\newblock {\em J. Cybernetics\/}, {\bf 3}, 58--72.

\bibitem[Bezdek(1975)Bezdek]{Bezdek75}
Bezdek, J.~C. (1975).
\newblock Mathematical models for systematics and taxonomy.
\newblock In G.~F. Estabrook, editor, {\em Proceedings of the 8th International
  Conference on Numerical Taxonomy\/}, San Francisco. Freeman.

\bibitem[Bezdek(1981)Bezdek]{Bezdek81}
Bezdek, J.~C. (1981).
\newblock {\em Pattern Recognition With Fuzzy Objective Function Algorithms\/}.
\newblock Plenum Press, New York.

\bibitem[Cho {\em et~al.}(1998)Cho, Campbell, Winzeler, Steinmetz, Conway,
  Wodicka, Wolfsberg, Gabrielian, Landsman, Lockhart, and Davis]{Cho98}
Cho, R.~J., Campbell, M.~J., Winzeler, E.~A., Steinmetz, L., Conway, A.,
  Wodicka, L., Wolfsberg, T.~G., Gabrielian, A.~E., Landsman, D., Lockhart,
  D.~J., and Davis, R.~W. (1998).
\newblock {{A} genome-wide transcriptional analysis of the mitotic cell cycle}.
\newblock {\em Mol. Cell\/}, {\bf 2}, 65--73.

\bibitem[Dave(1996)Dave]{Dave96}
Dave, R.~N. (1996).
\newblock {Validating fuzzy partition obtained through c-shells clustering}.
\newblock {\em Pattern Recogn. Lett.}, {\bf 17}, 613--623.

\bibitem[Demb\'el\'e and Kastner(2003)Demb\'el\'e and Kastner]{Dembele2003}
Demb\'el\'e, D. and Kastner, P. (2003).
\newblock {{F}uzzy {C}-means method for clustering microarray data}.
\newblock {\em Bioinformatics\/}, {\bf 19}, 973--980.

\bibitem[D{\"o}ring {\em et~al.}(2006)D{\"o}ring, Lesot, and Kruse]{Doring2006}
D{\"o}ring, C., Lesot, M.-J., and Kruse, R. (2006).
\newblock Data analysis with fuzzy clustering methods.
\newblock {\em Comput. Stat. Data An.}, {\bf 51}(1), 192--214.

\bibitem[Dunn(1973)Dunn]{Dunn73}
Dunn, J.~C. (1973).
\newblock A fuzzy relative of the isodata process and its use in detecting
  compact well-separated clusters.
\newblock {\em J. Cybernet.}, {\bf 3}, 32--57.

\bibitem[Eisen {\em et~al.}(1998)Eisen, Spellman, Brown, and Botstein]{Eisen98}
Eisen, M.~B., Spellman, P.~T., Brown, P.~O., and Botstein, D. (1998).
\newblock {{C}luster analysis and display of genome-wide expression patterns}.
\newblock {\em Proc. Natl. Acad. Sci. U.S.A.}, {\bf 95}, 14863--14868.

\bibitem[Fukuyama and Sugeno(1989)Fukuyama and Sugeno]{Fukuyama89}
Fukuyama, Y. and Sugeno, M. (1989).
\newblock A new method of choosing the number of clusters for the fuzzy c-means
  method.
\newblock {\em Proc. 5th Fuzzy Syst. Symp.}, page 247.

\bibitem[Futschik and Carlisle(2005)Futschik and Carlisle]{Futschik2005}
Futschik, M.~E. and Carlisle, B. (2005).
\newblock {{N}oise-robust soft clustering of gene expression time-course data}.
\newblock {\em J. Bioinform. Comput. Biol.}, {\bf 3}, 965--988.

\bibitem[Hanai {\em et~al.}(2006)Hanai, Hamada, and Okamoto]{Hanai2006}
Hanai, T., Hamada, H., and Okamoto, M. (2006).
\newblock {{A}pplication of bioinformatics for {D}{N}{A} microarray data to
  bioscience, bioengineering and medical fields}.
\newblock {\em J. Biosci. Bioeng.}, {\bf 101}, 377--384.

\bibitem[H\"oppner {\em et~al.}(1999)H\"oppner, Klawonn, Kruse, and
  Runkler]{Hoeppner99}
H\"oppner, F., Klawonn, F., Kruse, R., and Runkler, T. (1999).
\newblock {\em Fuzzy Cluster Analysis\/}.
\newblock John Wiley \& Sons, Inc., New York.

\bibitem[Horton and Nakai(1996)Horton and Nakai]{Horton96}
Horton, P. and Nakai, K. (1996).
\newblock {{A} probabilistic classification system for predicting the cellular
  localization sites of proteins}.
\newblock {\em Proc. Int. Conf. Intell. Syst. Mol. Biol.}, {\bf 4}, 109--115.

\bibitem[Iyer {\em et~al.}(1999)Iyer, Eisen, Ross, Schuler, Moore, Lee, Trent,
  Staudt, Hudson, Boguski, Lashkari, Shalon, Botstein, and Brown]{Iyer99}
Iyer, V.~R., Eisen, M.~B., Ross, D.~T., Schuler, G., Moore, T., Lee, J.~C.,
  Trent, J.~M., Staudt, L.~M., Hudson, J., Boguski, M.~S., Lashkari, D.,
  Shalon, D., Botstein, D., and Brown, P.~O. (1999).
\newblock {{T}he transcriptional program in the response of human fibroblasts
  to serum}.
\newblock {\em Science\/}, {\bf 283}, 83--87.

\bibitem[Krishnapuram and Freg(1992)Krishnapuram and Freg]{Krishnapuram92}
Krishnapuram, R. and Freg, C.-P. (1992).
\newblock Fitting an unknown number of lines and planes to image data through
  compatible cluster merging.
\newblock {\em Pattern Recogn.}, {\bf 25}, 385--400.

\bibitem[Nash {\em et~al.}(1994)Nash, Sellers, Talbot, Cawthorn, and
  Ford]{Nash94}
Nash, W.~J., Sellers, T.~L., Talbot, S.~R., Cawthorn, A.~J., and Ford, W.
  (1994).
\newblock {The Population Biology of Abalone (Haliotis species) in Tasmania. I.
  Blacklip Abalone (H. rubra) from the North Coast and Islands of Bass Strait}.
\newblock {\em Sea Fisheries Division Technical Report\/}, {\bf 48}.

\bibitem[Pal and Bezdek(1995)Pal and Bezdek]{Pal95}
Pal, N.~R. and Bezdek, J.~C. (1995).
\newblock On cluster validity for the fuzzy c--means model.
\newblock {\em Fuzzy Systems\/}, {\bf 3}, 370--379.

\bibitem[Pierce {\em et~al.}(2008)Pierce, Unwin, Evans, Griffiths, Carney,
  Zhang, Jaworska, Lee, Blinco, Okoniewski, Miller, Bitton, Spooncer, and
  Whetton]{Pierce2008}
Pierce, A., Unwin, R.~D., Evans, C.~A., Griffiths, S., Carney, L., Zhang, L.,
  Jaworska, E., Lee, C.~F., Blinco, D., Okoniewski, M.~J., Miller, C.~J.,
  Bitton, D.~A., Spooncer, E., and Whetton, A.~D. (2008).
\newblock {{E}ight-channel i{T}{R}{A}{Q} enables comparison of the activity of
  six leukemogenic tyrosine kinases}.
\newblock {\em Mol. Cell Proteomics\/}, {\bf 7}, 853--863.

\bibitem[Sigillito {\em et~al.}(1989)Sigillito, Wing, Hutton, and
  Baker]{Sigillito89}
Sigillito, V.~G., Wing, S.~P., Hutton, L.~V., and Baker, K.~B. (1989).
\newblock {Classification of radar returns from the ionosphere using neural
  networks}.
\newblock {\em John Hopkins APL Technical Digest\/}, {\bf 10}.

\bibitem[Tamayo {\em et~al.}(1999)Tamayo, Slonim, Mesirov, Zhu, Kitareewan,
  Dmitrovsky, Lander, and Golub]{Tamayo99}
Tamayo, P., Slonim, D., Mesirov, J., Zhu, Q., Kitareewan, S., Dmitrovsky, E.,
  Lander, E.~S., and Golub, T.~R. (1999).
\newblock {{I}nterpreting patterns of gene expression with self-organizing
  maps: methods and application to hematopoietic differentiation}.
\newblock {\em Proc. Natl. Acad. Sci. U.S.A.}, {\bf 96}, 2907--2912.

\bibitem[Tavazoie {\em et~al.}(1999)Tavazoie, Hughes, Campbell, Cho, and
  Church]{Tavazoie99}
Tavazoie, S., Hughes, J.~D., Campbell, M.~J., Cho, R.~J., and Church, G.~M.
  (1999).
\newblock {{S}ystematic determination of genetic network architecture}.
\newblock {\em Nat. Genet.}, {\bf 22}, 281--285.

\bibitem[Wolf-Yadlin {\em et~al.}(2007)Wolf-Yadlin, Hautaniemi, Lauffenburger,
  and White]{Wolf-Yadlin2007}
Wolf-Yadlin, A., Hautaniemi, S., Lauffenburger, D.~A., and White, F.~M. (2007).
\newblock {{M}ultiple reaction monitoring for robust quantitative proteomic
  analysis of cellular signaling networks}.
\newblock {\em Proc. Natl. Acad. Sci. U.S.A.}, {\bf 104}, 5860--5865.

\bibitem[Wu {\em et~al.}(2005)Wu, Yu, and Yang]{Wu2005}
Wu, K.-L., Yu, J., and Yang, M.-S. (2005).
\newblock A novel fuzzy clustering algorithm based on a fuzzy scatter matrix
  with optimality tests.
\newblock {\em Pattern Recogn. Lett.}, {\bf 26}(5), 639--652.

\bibitem[Xie and Beni(1991)Xie and Beni]{Xie91}
Xie, X.~L. and Beni, G. (1991).
\newblock A validity measure for fuzzy clustering.
\newblock {\em IEEE Trans. Pattern Anal. Mach. Intell.}, {\bf 13}(8), 841--847.

\bibitem[Zadeh(1965)Zadeh]{Zadeh65}
Zadeh, L.~A. (1965).
\newblock Fuzzy sets.
\newblock {\em Inf. Control\/}, {\bf 8}(3), 338--353.

\end{thebibliography}

\end{document}